%% file: main.tex
\documentclass[amsmath,amssymb,aps,prl,reprint,floatfix,superscriptaddress]{revtex4-1}

\usepackage[utf8]{inputenc}
\usepackage{times}
\usepackage{changes}
\usepackage{color}
\usepackage{amsfonts}
\usepackage{amsmath,amssymb}
\usepackage[english]{babel}
\usepackage{subfigure}
\usepackage{graphicx}
\usepackage{epsf} 
\usepackage{csquotes}
\usepackage{hyperref}
\usepackage{gensymb}

\usepackage{wrapfig}
\newcommand{\ket}[1]{|#1\rangle}

\begin{document}

\title{A trapped ion quantum computer with robust entangling gates and quantum coherent feedback}

\author{Tom Manovitz}
\email{tom.manovitz@weizmann.ac.il} \thanks{Equal contribution}
\author{Yotam Shapira} \thanks{Equal contribution}
\author{Lior Gazit}
\author{Nitzan Akerman}
\author{Roee Ozeri}
\affiliation{%
Department of Physics of Complex Systems and AMOS, Weizmann Institute of Science, Rehovot 7610001, Israel
}%

\date{September 2021}

\begin{abstract}
    \input{abstract}
\end{abstract}

\maketitle

\section{Introduction}

\input{Sections/introduction}

\section{System overview}
\input{Sections/tools}

\section{Individual qubit addressing and detection} \label{individual}
\input{Sections/addressing}

\section{Robust Entangling Gates}
\input{Sections/multiqubit}

\section{Coherent feedback}
\input{Sections/feedback}

\section{Conclusion}
\input{Sections/conclusion}

\bibliography
\bibliographystyle{apsrev4-1}

\end{document}

%% file: abstract.tex
Quantum computers are expected to achieve a significant speed-up over classical computers in solving a range of computational problems. Chains of ions held in a linear Paul trap are a promising platform for constructing such quantum computers, due to their long coherence times and high quality of control. Here we report on the construction of a small, five-qubit, universal quantum computer using $^{88}\text{Sr}^{+}$ ions in an RF trap. All basic operations, including initialization, quantum logic operations, and readout, are performed with high fidelity. Selective two-qubit and single-qubit gates, implemented using a narrow linewidth laser, comprise a universal gate set, allowing realization of any unitary on the quantum register. We review the main experimental tools, and describe in detail unique aspects of the computer: the use of robust entangling gates and the development of a quantum coherent feedback system through EMCCD camera acquisition. The latter is necessary for carrying out quantum error correction protocols in future experiments.

%% file: Sections/introduction.tex
Ion traps are among the leading quantum computing platforms explored today due to their strong isolation from the environment and  naturally long coherence times. Trapped-ion quantum systems offer extremely accurate and reliable single qubit and many qubit operations; high fidelity state preparation and measurement; and high-connectivity between qubits, supported by the long-range Coulomb interaction between ions \cite{Ozeri2011Toolbox}.

Several approaches are currently pursued in order to scale-up trapped-ion quantum computing to a macroscopic number of qubits. These include shuttling of ions between different trapping regions in a CCD-like array of traps \cite{Kielpinski2002,Wan2019,Pino_2021}, and the heralded entanglement of ions in different traps using Hong-Ou-Mandel interference and detection of photons that were scattered by these ions \cite{monroe2014large,hucul2015modular}. While experimental progress along these two lines is encouraging, large scale trapped-ion quantum computing is yet to be demonstrated. 

In the intermediate term, trapped-ion quantum computing is performed mostly using registers of few-to-several tens of trapped-ion qubits in a single Coulomb crystal. In these systems, different ions in the register are often individually addressed with focused laser beams. Entanglement is generated using the motional modes of the ion crystal as spin-dependent force mediators between ions. The collective nature of the crystal motional modes leads to all-to-all connectivity between ion qubits. This high connectivity was shown to be an advantage in comparisons with other qubit technologies \cite{Linke3305}.  State detection of different qubits in the register is performed in parallel by state-selective fluorescence and the imaging of these photons on a sensitive CCD camera or an array of photon-detectors. Only in very few such demonstrations the ability to perform mid-circuit measurement and conditional coherent feedback was shown \cite{egan2020fault}. 

Several small trapped-ion quantum registers were demonstrated and benchmarked in the last few years. Notable examples include Ca$^+$ registers of optical qubits at the University of Innsbruck and Alpine Quantum Computing \cite{Schindler2013,IQT2021}; Yb$^+$ registers by JQI, IonQ \cite{Debnath2016,IonQ2019} and the Tsinghua group \cite{Tsinghua2019}; and shuttling-based trapped ion registers by NIST,  ETH, Honeywell and Mainz \cite{NIST2009, ETH2018, Honeywell2021, Mainz2021}.

Mid-circuit measurements, and the ability to coherently operate on qubits conditioned on the measurement outcome, is an integral part of many-quantum computing protocols, most notably quantum error-correction. Mid-circuit measurement and coherent feedback in multi-ion crystals present several challenges. Firstly, in state-selective fluorescence it is difficult to prevent neighbouring, quantum-information carrying, ions from scattering photons and thus losing coherence. In recent demonstrations this difficulty was mitigated by hiding logic ions through encoding them on superpositions in a meta-stable optical level \cite{Schindler2013}, by using different atomic species as readout and logic ions \cite{NIST2009, ETH2018}, and by shuttling ions to separate measurement regions \cite{NIST2009, Honeywell2021, Mainz2021}. Secondly, coherent feedback requires the extraction of the measurement outcome to a classical control system in times shorter than the system's coherence time. Multi-ion crystals are often read-out by CCD cameras that spatially resolve the different ions. CCD readout, however, has a non-negligible constant time-delay determined by the transfer time of charge through the CCD. In  previous demonstrations of trapped-ions mid-circuit measurement followed by coherent feedback, the readout was performed using a fast photo-detectors rather than a CCD camera.

Another challenge of quantum computing in multi-ion crystals is preserving their high fidelity when applied to ion-pairs selected from a multi-ion crystal. The higher effective mass of the crystal and the presence of more normal modes renders entanglement gates more susceptible to noise. Recently, robust entanglement gates were demonstrated using coherent control and dynamic decoupling techniques in two-ion crystals \cite{Shapira2018,webb2018resilient,zarantonello2019robust,Manovitz2017}. The adaptation of such robust gates to multi-ion crystals is therefore a next important step. A recent publication on an 11-qubit quantum computer reported the implementation of entanglement gates that are robust against trap-frequency errors \cite{IonQ2019}. 


Here we describe a 5-qubit quantum computer built at the Weizmann Institute of Science. Our computer is based on a single crystal of Sr$^+$ ions. Qubits are encoded either on optically separated orbitals or in ground-state spin states. We demonstrate mid-circuit measurement followed by coherent feedback using an ion resolving imaging capability on a CCD camera. Furthermore, we apply two-qubit gates on all ion pairs, using the crystal center-of-mass mode of motion. Our gates were engineered to be robust against gate timing errors and off-resonance carrier coupling. Such small-scale quantum computing modules are expected to be the building blocks of future large-scale trapped ion computers and therefore the improvement of their performance is important.

%% file: Sections/tools.tex
\begin{figure}
\centering
	\includegraphics[width=1\columnwidth]{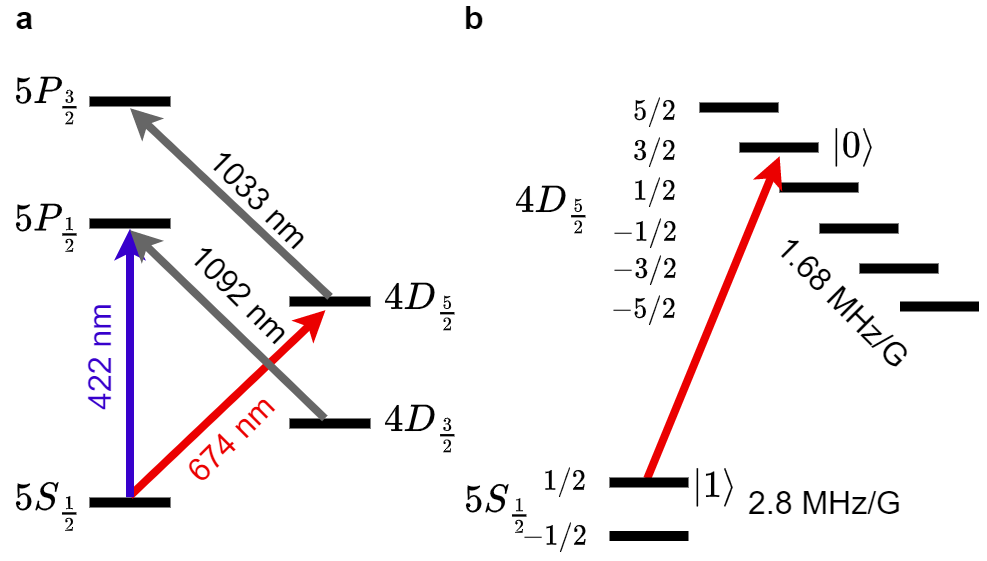}
	\caption[$^{88}\text{Sr}^+$ level structure.]{$^{88}\text{Sr}^+$ level structure. (a) A 422 nm laser excites the dipole transition from the ground state to the short lived (8 ns) $5P_\frac{1}{2}$ state, used for Doppler cooling, EIT cooling, optical pumping, and state-selective fluorescence state detection. A 674 narrow-linewidth laser drives the narrow, dipole-forbidden transition between the ground state and the long-lived (390 ms) excited state $4D_\frac{5}{2}$, used for sideband cooling and coherent control of the optical qubit. 1092 and 1033 nm lasers are used to repump the $4D_\frac{3}{2}$ and $4D_\frac{5}{2}$ levels, correspondingly. (b) A magnetic field Zeeman splits all orbitals, including the ground and $4D_\frac{5}{2}$ orbitals which are used to encode an optical qubit. Typically, the qubit is encoded on $\ket{1} = \ket{5S_\frac{1}{2},\frac{1}{2}}$ $\ket{0} = \ket{4D_\frac{5}{2},\frac{3}{2}}$, as shown.}
	\label{levels}
\end{figure}

Our quantum computer is based on a chain of $^{88}\text{Sr}^+$ ions held in a linear Paul trap. The alkaline earth $^{88}\text{Sr}^+$ possesses a single valence electron, resulting in a simple level structure with optical transitions at convenient visible or near-infrared wavelengths (422, 674, 1092 and 1033 nm), easily generated using diodes lasers. The relevant level structure for $^{88}\text{Sr}^+$ includes the ground state $5S_{\frac{1}{2}}$; two long-lived excited states $4D_{\frac{3}{2}}$ (435 ms lifetime) and $4D_{\frac{5}{2}}$ (390 ms); and two short-lived excited states, $5P_{\frac{1}{2}}$ (8 ns) and $5P_{\frac{3}{2}}$ (8 ns). A diagram of the these levels is shown in Fig. \ref{levels}.

Two electromagnetic coils in a Helmholtz configuration produce a constant  magnetic field of $3-5$ G in the vicinity of the ions, defining a quantization axis and Zeeman splitting all levels. A qubit can be encoded on the ion in two ways: A Zeeman qubit is encoded using the two Zeeman-split (2.802 MHz/G) spin levels of the ground state $5S_{\frac{1}{2}}$ manifold, while an optical qubit is encoded by pairing one of the states in this manifold with one of the states of the $4D_{\frac{5}{2}}$ orbital manifold (1.68 MHz/G). We primarily make use of the optical qubit. In order to store and protect qubit information during measurement, we encode qubits entirely in the $4D_{\frac{5}{2}}$ orbital manifold. Moreover, in order to selectively hide qubits from interaction with the addressing laser, we may encode these qubits on varying pairs of states in the $5S_{\frac{1}{2}}$ and $4D_{\frac{5}{2}}$ manifolds.

The ions are spatially confined to a one dimensional chain by a linear Paul trap. A detailed description of our trap and vacuum system may be found in \cite{Akerman2011,Akerman:2012}. The trap RF is  at approximately 21 MHz; secular (harmonic) trapping frequencies are approximately 1-1.5 MHz along the trap axis, corresponding to typical inter-ion spacing of about 3-4 micrometers; and 3-4 MHz along the radial axes. In order to load ions into the trap, neutral strontium vapor, produced by heating nearby ovens, is photo-ionized in the trapping region through a two-photon process: the $5s^{2}$ $^{1}S_0$ ground state is excited to the $5s5p$ $^1P_1$ state and then to the self-ionizing $5p^2$ $^1D_2$ state, using corresponding 461 nm and 405 nm lasers. 

Optical control of the ions is realized using lasers at wavelengths of 422 $(5S_\frac{1}{2}\leftrightarrow 5P_\frac{1}{2})$, 674 $(5S_\frac{1}{2}\leftrightarrow 4D_\frac{5}{2})$, 1033 $(4D_\frac{5}{2}\leftrightarrow 5P_\frac{3}{2})$ and 1092 nm $(4D_\frac{3}{2}\leftrightarrow 5P_\frac{1}{2})$. Lasers enter the vacuum chamber through fused-silica windows on the sides or the top of the chamber. All lasers are generated by diodes in an ECDL configuration and are locked to stable external references. Of particular importance is the 674 nm laser, which is used to coherently control the optical qubit. This laser is first locked to a ultra-low expansion (ULE) Fabry-Perot cavity with high servo bandwidth; in order to filter out the lock servo bumps, the cavity transmission is amplified by a slave diode, and the output is then locked to an additional temperature-stabilized ULE Fabry-Perot cavity, generating narrow-linewidth light \cite{Peleg2019}.

The ion chain is typically Doppler cooled to a temperature of a few mK using the 422 nm laser, together with a 1092 nm laser to repump the long-lived $4D_\frac{3}{2}$ orbital. The chain may then be further cooled to the ground state of several, or all, motional modes using sideband and EIT cooling. Sideband cooling is used to cool axial motional modes, and is performed by tuning the 674 nm laser near the red-sideband resonance of the relevant mode, together with the 1033 nm repump and 422 nm optical pumping lasers. Several modes may be cooled in parallel by driving the 674 nm laser using multiple frequencies. EIT cooling is used to cool all radial modes in parallel. A 422 nm $\sigma^+$-polarized pump and $\pi$-polarized probe isolate an effective $\Lambda$-type 3-level system with a $k$-vector that is perpendicular to the trap axis and equally overlaps both radial trapping axes. Blue detuned from the transition, the pump creates a narrow asymmetric (Fano) resonance when addressed by the probe beam, enabling efficient cooling with low background scattering \cite{lechner2016eit}. 

Qubit initialization and measurement are realized using the 422 nm laser, together with the 1092 repump laser to prevent pumping into the long-lived $4D_{\frac{3}{2}}$ orbital. The qubits are prepared in the qubit ground state $\ket{1} = \ket{5S_{\frac{1}{2},+\frac{1}{2}}}$ by optical pumping using a $\sigma^+$ polarized 422 nm beam. This process prepares $\ket{1}$ with an error smaller than $10^{-3}$ within $10$ $\mu$s. Strong measurement of the optical qubit is achieved by state-selective fluorescence: ions are illuminated with the   light, which scatters off ions occupying $\ket{1}$ but not off those occupying the qubit excited state $\ket{0} = \ket{4D_{\frac{5}{2},+\frac{3}{2}}}$ (or any other state in the $\ket{4D_{\frac{5}{2}}}$ manifold). The scattered fluorescence is imaged onto an electron-multiplying CCD (EMCCD) camera, from which the  state is determined for each ion separately. Fluorescence is collected for 1 ms, giving rise to an error of $2.6\times 10^{-3}$ due to the finite lifetime of $\ket{0}$. 

Coherent operations are performed with either the 674 nm laser (optical qubit) or an RF magnetic field (Zeeman qubit). The latter is induced by running RF current through an electrode which is 2 mm away from the ions, and is used to drive transitions between Zeeman states within the $5S_{\frac{1}{2}}$ and $4D_{\frac{5}{2}}$ manifolds. A narrow-linewidth \cite{Peleg2019} 674 nm laser, illuminating the ions through a global and an individual addressing path, is used to drive coherent optical qubit operations. The global path overlaps all ions in the chain with approximately uniform intensity. By resonantly addressing the ions, this path can be used to apply global unitary rotations of the form,
\begin{equation}
    U_{G}(\phi,\theta) = \exp{\left(i\frac{\theta}{2}\sum_i\sigma^\phi_i\right)}.
\end{equation}
Here $\sigma^\phi_i = \cos{(\phi)}X_i+\sin{(\phi)}Y_i$, while $X_i$ and $Y_i$ are the standard Pauli operators acting on the optical qubit subspace of ion $i$. Using the M{\o}lmer-S{\o}rensen interaction, the global path can also be used to generate global entangling operations of the form,
\begin{equation}
    U_{MS}(\phi,\theta) = \exp{\left(i\frac{\theta}{2}\sum_{i<j}\sigma^\phi_i\sigma^\phi_j\right)}.
\end{equation}
The individual addressing path is tightly focused onto a single ion at a time, and can be quickly steered between ions using a pair of acousto-optic deflectors (AODs) in an XY (perpendicular) configuration. For this path, the laser is tuned off resonance to generate a light shift, and can therefore realize the  operations,
\begin{equation}
    U_i(\theta) = \exp{\left(i\frac{\theta}{2}Z_i\right)}.
\end{equation}
Here $Z_i$ is the according Pauli operator acting on the optical qubit subspace of ion $i$. Together, these $U_{G}(\phi,\theta)$, $U_{MS}(\phi,\theta)$ and $U_i(\theta)$ operations constitute a universal gate set on the optical qubit register. Below we describe our implementation of entangling and individual addressing operations in more detail.


%% file: Sections/addressing.tex
Universal quantum computation requires the ability to both operate on, and measure, the state of any specific qubit in the register. In ion traps, the inter-ion spacing of a few micrometers \cite{james2000quantum} is larger than optical wavelengths, implying that one can use focused laser beams for resolving the position of each ion in the chain \cite{Schindler2013,Debnath2016,pogorelov2021compact}. However, while larger than optical wavelengths, inter-ion spacing is not \emph{much} larger, meaning high numerical apertures are necessary and no more than minor aberrations can be tolerated for individual addressing to work with sufficiently low cross-talk error. Other approaches to individual addressing include spectrally resolving ions \cite{mintert2001ion,johanning2009individual,timoney2011quantum}, coupling ions to spatially varying oscillating fields \cite{navon2013addressing}, or shuttling them to separate trapping regions \cite{Kielpinski2002,Wan2019,Pino_2021}).

We performed both qubit-resolving state measurements as well as individual addressing for single qubit rotations using the same high-resolution optical system. Qubits were measured by collecting state-selective fluorescence on  the $5S_\frac{1}{2}\leftrightarrow5P_\frac{1}{2}$ transition at 422 nm, which was imaged onto an EMCCD camera (Andor iXon Ultra 897). We individually addressed qubits by shining 674 nm light, near-resonant with the optical $5S_\frac{1}{2}\leftrightarrow4D_\frac{5}{2}$ transition, in a tightly focused beam that spatially overlaps only a single ion at a time. The laser beam is quickly steered between ions using two orthogonal acousto-optic deflectors (AODs).

The imaging system objective (LENS-optics) has an effective focal length of 30 mm and a NA of 0.34, and is designed to correct for chromatic aberrations of its lenses as well as the vacuum chamber fused-silica windows at both relevant wavelengths (422 nm for detection and 674 nm for addressing). The objective is mounted on a Newport xyz translation stage for positioning, which is itself connected to a custom-made holder that is tightly anchored to the vacuum chamber flange. This allows the objective to be positioned a few mm away from the vacuum chamber window, which is sunken into the chamber and enables an effective working distance of 30 mm from the ions. Anchoring the objective to the chamber minimizes vibrations relative to the ion position, suppressing beam pointing noise, which can be significant for tightly focused beams. The objective is designed to work with the ions in its focal-plane using a $\infty-f$ configuration. We slightly shift the objective away from this working point so that it directly images the ions on the EMCCD at a distance of approximately $120$ cm. This enables imaging the ions without adding any additional optical elements, thus reducing aberrations. 

Apart from the camera, all other optical elements are positioned on an optical breadboard mounted directly above the trap by five aluminum legs. Diagonal brace beams are added where possible to mitigate table vibrations. A 1.5-inch diameter hole is bored through the table, directly above the objective and the trap. A pair of mirrors change the optical beam path from the horizontal path on the breadboard onto the vertical path descending into the trap, and vice versa. Near these mirrors, a longpass dichroic mirror separates the detection and addressing beam paths \cite{Manovitz:2016}.

\subsection{Individual addressing}

The narrow linewidth 674 nm light is passed through a double-pass AOD and a single pass AOM that provide amplitude and phase control. The light is then delivered with an optical fiber to the optical breadboard on which it passes through a pair of Isomet OAD948-633 AODs, held in an XY configuration. The zero-order diffraction of the second AOD is picked up by a detector; the signal from this detector is sent to a PID that controls an attenuator on the RF for the single pass AOM in the 674 nm individual addressing line, and is used to stabilize the laser power. The diffracted light passes through a pair of lenses, followed by the dichroic mirror, and is focused onto the ions by the objective lens. 

The AODs were chosen to support the contradicting demands of high resolution and short switching time. The switching time $\tau_s$ is determined by the propagation time of the acoustic wave through the laser waist of size $d$ on the AOD: $\tau_s = d/v$, where $v$ is the acoustic velocity. The resolution $N$, defined as the number of resolved optical spots that the deflector can scan, is given by $N=\tau_s \Delta f$, where $\Delta f$ is the frequency bandwidth of the AOD. Hence, resolution can be increased at the cost of longer switching times. The Isomet AODs provide a bandwidth $\Delta f = 45$ MHz and an acoustic velocity of $v = 650$ m/s. Hence, a spot size of 1.1 mm defines switching time $\tau \approx 1.7 \mu s$ and resolution $N\approx 77$. 

The AODs are set in an XY configuration so that the beam can be scanned across a plane, rather than a line, enabling precise positioning of the beam focus on each ion. The orientation relative to the trap is such that the ion chain corresponds to a $45^\circ$ angle with respect to axes defined the XY configuration, and opposite frequencies are added to the two AODs in order to scan across the ion chain. Such a configuration simplifies control, as the 674 nm frequency does not need to be corrected when the focus position is changed from one ion to another. Taking the XY configuration into account, the resolution is effectively increased to $N\approx 108$. These parameters are sufficient, as they support low time overhead compared to single qubit gates $(\sim 10 \mu s)$, and can easily accommodate a large ion chain $(<30 \mu m)$.

\begin{figure*}
\centering
	\includegraphics[width=1\linewidth]{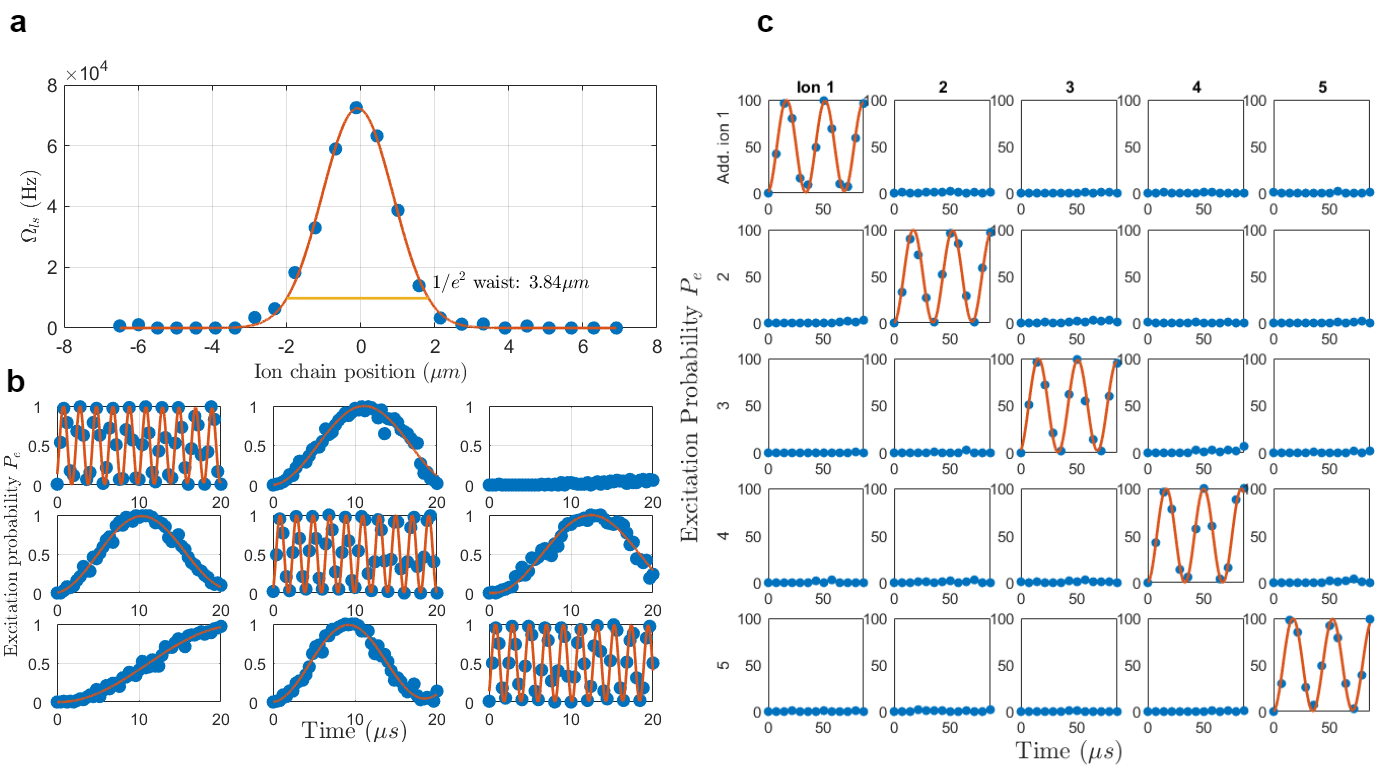}
	\caption[Individual addressing intensity profile on ion.]{a) Beam intensity profile of the individual addressing beam along the ion chain axis, determined by measuring the position-dependent $\Omega_{ls}$, which is proportional to the intensity. We measure a $1/e^2$ waist diameter of $2w_0=3.84 \mu m$. b) In order to measure cross-talk even for low values, we resonantly address the ions using the individual addressing path. Cross-talk Rabi frequency ratio is approximately $10^{-1}$ for $3.71 \mu m$ inter-ion distance. c) An example of single qubit Rabi oscillations using individual addressing in the light shift configuration. Here we use a five-ion chain, with a minimal inter-ion distance of $2.83 \mu m$.}
	\label{crosstalk}
\end{figure*} 

The individual addressing system is used to address the ions off-resonantly, generating a light shift with effective Rabi frequency $\Omega_{ls} = \frac{\Omega^2_{res}}{\Delta}$, where $\Omega_{res}$ is the resonant Rabi frequency and $\Delta$ is the detuning from resonance. Typical parameters are $\Omega_{res}= 500$ kHz and $\Delta = 5$ MHz, implying an effective light shift of $\Omega_{ls}=50$ kHz. Limiting the individual path to $\sigma_z$ rotations via off resonant illumination has significant advantages as compared with resonant coupling. Coherent resonant coupling from both global and individual paths require maintaining phase coherence between the two, which is challenging due to the very different optical trajectories. Furthermore, off-resonant addressing minimizes crosstalk by effectively decreasing the individual path spot size. While resonant coupling strength is determined by the Rabi frequency $\Omega_{res}$, off-resonant coupling is proportional to $\Omega_{res}^2$; hence the effective beam cross section depends on the amplitude squared rather than the amplitude and leads to a $\sqrt{2}$ narrower cross-section. $\sigma_z$ rotations in conjunction with $\sigma^\phi$ global rotations can generate any single-qubit unitary.  


We characterize the beam profile by measuring the lights-hift on the ion, which is proportional to the laser intensity, as a function of AOD frequency, or correspondingly, beam position. The results are shown in Fig. \ref{crosstalk}. A useful measure of the spot size is the beam waist diameter $2w_0$, which is the $1/e^2$ intensity diameter of a Gaussian beam. Making use of the full aperture of our 0.34 NA objective, for a diffraction limited beam, one would expect a beam waist diameter $2w_0 = \frac{2\lambda}{\pi NA} \approx 1.3$ $\mu m$. However, we measure a significantly larger beam waist diameter of $2w_0 =3.84$ $\mu m$. The discrepancy is a result of an effectively smaller NA and aberrations. Use of the entire aperture of the objective would clip the outer tail of the Gaussian profile; this would imply a focus on the ions that is a convolution of a Gaussian and an Airy beam profile. The latter introduces intensity oscillations that decay with distance from the beam center $r$ as $1/r^2$, which may then generate non-negligible cross-talk between ions. To avoid significant Airy-like features, the beam waist diameter on the objective is set to be smaller than the objective aperture, implying an effective NA of approximately $NA\approx 0.19$ which, for a diffraction-limited spot, would result in a waist diameter $2.3 \mu m$. In practice, we achieve a diameter that is $\sim1.7$ times larger due to aberrations, which are captured by a beam quality parameter $M^2\approx 1.7$.

As long as the distance between ions is larger than the beam waist radius of $w_0=1.92 \mu m$, which is always true for our experiments, the cross-talk is determined by the beam tail. Perfectly Gaussian beams decay strongly, so in practice tails are dominated by small aberrations, such as weak Airy contributions from the aperture cutoff, or minor astigmatism and coma from imperfect alignment. Hence, the beam waist is not necessarily a reliable parameter for evaluating cross-talk. The cross-talk parameter of interest is $R_{ls}= \max_{i,j}{\frac{\Omega^{i,j}_{ls}}{\Omega^{i,i}_{ls}}}$, where $\Omega_{ls}^{i,j}$ is the light-shift induced on ion $j$ when attempting to address ion $i$. $R_{ls}$ can be used to give an upper bound on errors generated by cross-talk per single-qubit operation: $\epsilon_{crosstalk}\leq 2R_{ls}^2$ (under the assumption that nearest-neighbour cross-talk is the main source of error). As the $R_{ls}$ is quite low, it is difficult to properly measure directly. Instead, we measure the resonant cross-talk parameter $R_{res}= \max_{i,j}{\frac{\Omega^{i,j}_{res}}{\Omega^{i,i}_{res}}}$, from which one can straightforwardly derive $R_{ls}=R_{res}^2$. To do so, we trap a 3-ion chain with inter ion distances $3.7 \mu m$ (a typical inter-ion distance for small chains in our trap),  drive resonant Rabi oscillations while attempting to address each ion, and measure the Rabi frequency on all ions simultaneously. The results are shown in Fig. \ref{crosstalk}. For an inter-ion distance of 3.7 $\mu m$s, we measure $R_{res}=0.12(1)$, implying $R_{ls}=0.014(2)$ and $\epsilon_{crosstalk}\leq 3.9(5)\times10^{-4}$. Hence, for such inter-ion distances, the error induced by cross-talk is significantly lower than other error sources. An example of single qubit Rabi rotations in a five-qubit chain, induced by light-shift individual addressing, is shown in Fig. \ref{crosstalk} (c), with no discernible cross talk. 

The error incurred by an individual addressing operation on one of the ions is not limited to cross-talk; indeed, we find that it is dominated by other terms. We measured the fidelity of an individual addressing $\sigma^z$ $\pi$ angle rotation. This was achieved by initializing the 3 ions in a $\ket{111}$ state, applying a global $\sigma^x$ $\pi/2$ pulse, followed by an individual addressing $\sigma^z$ $\pi$ pulse on one of the  ions, and then a global $\sigma^x$ $-\pi/2$ pulse (applied by varying the phase of the global 674 nm laser by $\pi$). The desired result is a chain in which only one of the qubits is in the excited state, e.g. the $\ket{101}$ state. After correcting for state preparation and measurement (SPAM) errors due to the global shelving pulses, optical pumping initialization, and measurement fidelity, we arrive at a fidelity of $0.9964(10)$ for a $\sigma^z$ $\pi$ pulse. We also find that while the target ion undergoes an operation, other ions suffer an error of $0.001-0.002$ per ion; this may be due to background $\sigma^z$ noise.

\subsection{Detection} 

Qubit state detection is performed through state-selective fluorescence scattering of the 422 nm laser on the $5S_\frac{1}{2}\leftrightarrow 5P_\frac{1}{2}$ transition: fluorescence is observed only if the electron collapses to the $5S_\frac{1}{2}$ state. The scattered fluorescence is picked up by the LENS-Optics 0.34 NA objective, which focuses the images of the ions directly onto an Andor iXon Ultra 897 electron multiplying charge coupled-device (EMCCD) camera, placed approximately 1230 mm away from the objective. This results in a magnification of about $\times41$, measured by comparing distance between ions on camera to calculated values \cite{james2000quantum}. Apart from the objective itself and aligning mirrors, the only optical element in the path of fluorescence light is a dove prism; the prism is used to rotate the chain image such that it is aligned horizontally on the EMCCD, and is parallel with the pixel axis, allowing for a smaller pixel region of interest (ROI) and hence faster readout. The camera is placed on a custom made holder, physically separated from any other element in the system, in order to suppress detrimental acoustic vibrations caused by the camera cooling fans.

The main advantage of camera detection is the spatial resolution that it provides, allowing for simple differentiation between ions. Beyond the optical resolution, the iXon Ultra has high quantum efficiency; high readout speeds for a CCD camera; thermoelectrically cools the CCD array to $-90^\circ$ C, substantially mitigating thermal detection noise; and possesses an electron-multiplying capability that can increase gain to $\times500$. These attributes are important advantages for ion detection, giving good signal-to-noise ratios (S/N) for short detection times. However, the iXon Ultra uses a masked CCD storage area (designed to improve duty cycle); transferring the signal through the additional CCD frame storage area implies a readout time overhead of approximately $300-400 \mu s$. 


Typically, we define a rectangular region of interest (ROI) on the camera of approximately $6\times20-40$ pixels, where the latter depends on the number of ions in the chain. Vertically smaller ROIs result in lower readout times for the camera, and are thus beneficial. Readout times can be further minimized by placing the ion chain image in the corner of the pixel array that is closest to the analog-to-digital-converter (ADC). We achieve typical readout times (not including the detection exposure time itself) of 600-700 $\mu s$ including the frame storage overhead.


During a standard calibration procedure, a set of 1500 images of the ions fluorescing, and 1500 images of the ions not fluorescing, are acquired. Then, using these images, a set of pixels are allocated to each ion. For each ion, the distribution of the signal sum on the corresponding pixels are compared for the fluorescing and non fluorescing case, and an intermediate value is chosen as a threshold discriminator. After calibration, for every measurement and for every ion, the pixel sum is compared to this threshold in order to differentiate the fluorescing state (implying $\ket{1}=5S_\frac{1}{2}$) from a non fluorescing state (implying $\ket{0}=4D_\frac{5}{2}$).

For exposure times of 1 millisecond, we find that detection error is dominated by spontaneous emission of the $4D_{5/2}$ orbital, giving a detection error of $1-e^{-1/390} \approx 2.6\times10^{-3}$. We perform $2\times10^4$ detection measurements, where the ion state is predetermined by the application; or lack thereof; of the 1092 nm repump laser. When this laser is turned off, spontaneous emission errors are practically undetectable; if an emission event occurs, the electron is quickly pumped back into the non-fluorescing $4D_\frac{3}{2}$ state and remains there until another emission event or the end of the measurement. With spontaneous emission suppressed, we find no errors for these $2\times10^4$ measurements, implying with high confidence an error rate lower than $<10^{-3}$ from other sources. Thus we conclude the spontaneous emission error is dominant.

\begin{figure}
\centering
	\includegraphics[width=1\linewidth]{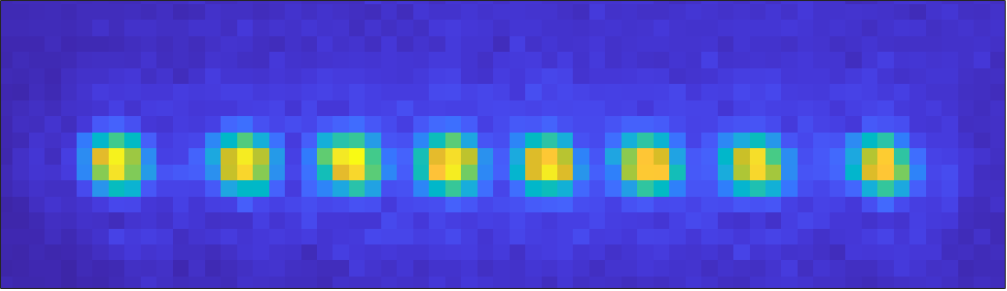}
	\caption[Image of 8 ions.]{An image of an 8-ion chain on the iXon Ultra camera. Here the minimal inter-ion distance is about $2.3 \mu m$. The ion chain is aligned along the pixel axis using a dove prism set before the camera.}
	\label{8ion}
\end{figure} 

For a series of acquisitions, each image is read via ADC and saved in the camera's internal buffer. The buffer is read out at the end of the series by a USB, where the measurement results are determined retroactively through the simple image analysis described above. In certain scenarios (e.g. QEC) determining the measurement result retroactively is not good enough, and a real-time decision must be made. We achieve this as well with the Andor iXon camera, as described in a succeeding section detailing our real time quantum coherent feedback module.

%% file: Sections/multiqubit.tex
Entangling gates are a key component in any realization of a quantum computer. They are typically the most challenging gates of the universal gate set and in ion quantum processors are often substantially slower and less robust than single qubit operations. The most prevalent method to entangle qubits in an ion trap quantum processor is by using the M{\o}lmer-S{\o}rensen (MS) technique \cite{sorensen1999quantum,sorensen2000entanglement} in which a spin-dependent force is exerted on a collective normal-mode of motion of the ion chain, resulting in an effective qubit-qubit interaction.

MS gates are very successful in two-ion chains \cite{gaebler2016high,ballance2016high,srinivas2021high,clark2021high}. However, in long ion-chains, the gates have an increased sensitivity to even small implementation imperfections. Here we utilize an entangling gate technique  that employs coherent control in order to decrease the sensitivity of the gate operation to experimental imperfections \cite{Shapira2018,webb2018resilient}. This technique is an extension of the MS gate and is beneficial for scaling up the ion qubit register. We demonstrate our entangling scheme on pairs of ions in a five ion chain, and in addition as an all-to-all gate in a four ion chain.

Our method is based on driving the gate with a multi-tone laser that is tuned on resonance with the optical qubit transition along with a sequence of amplitude modulations at frequencies $\omega_j=\nu+n_j\xi$, where $\nu$ is the frequency of the normal-mode of motion used to mediate the interaction, $n_j$ is an integer and $\xi$ is a detuning, such that the gate time is $T=2\pi/\xi$. One may write the interaction Hamiltonian due to this driving field as,

\begin{equation}
    V=\frac{1}{2}\Omega\sum_i\sigma^x_i\cos(kx_i-\omega_{eg}t)\sum_j r_j\cos\left(\omega_j t\right). 
\end{equation}
Here $\Omega$ is the resonant Rabi frequency, $\sigma^x_i$ is the Pauli operator on the optical qubit subspace of ion $i$, $k$ is the optical wavenumber, $x_i$ is the position of ion $i$, $\omega_{eg}$ is the resonant frequency for the optical qubit transition,and  $r_j$ represents the magnitude of the amplitude modulation at frequency $\omega_j$. The regular MS gate is formed by the special choice of a single modulation frequency with $n=1, r=1$, $\xi = 2\eta\Omega$ and $\eta=k x_0$, where $x_0$ is the ground state extent of the relevant motional mode.

\begin{figure}
\centering
	\includegraphics[width=1\linewidth]{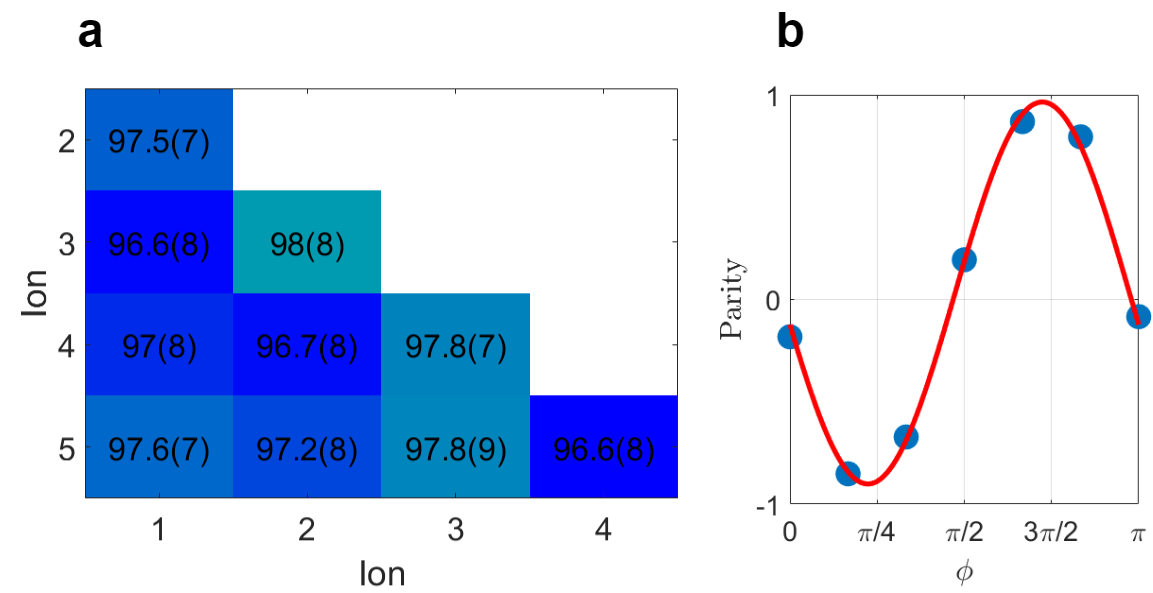}
	\caption[Entangling gate fidelity matrix.]{(a) Fidelities for generating a two-qubit maximally entangled state by applying an entangling gate for every pair of qubits on a 5-ion chain. The values are SPAM-corrected, with an average fidelity of 0.973(2). The entangling gate is a cardioid (1,3) M{\o}lmer-S{\o}rensen, applied with an echo pulse midway through the gate. (b) An example parity measurement on a pair of qubits in the chain.}
	\label{fidmatrix}
\end{figure} 

A derivation similar to that found in \cite{Shapira2018} shows that the fidelity of a two ion entangling gate in an $N$ ion chain is given by,
\begin{equation}
\begin{split}
    F_{2,N}  = & \frac{1}{8}\bigg[4e^{-\frac{F^{2}+G^{2}}{2}\left(\bar{n}+\frac{1}{2}\right)}\cos\left(N\left(A+\frac{FG}{2}-\frac{\pi}{2}\right)\right)\\
    & \cdot\cos\left(A+\frac{FG}{2}-\frac{\pi}{2}\right)+e^{-2\left(F^{2}+G^{2}\right)\left(\bar{n}+\frac{1}{2}\right)} \\
    & \cdot\cos\left(2N\left(A+\frac{FG}{2}-\frac{\pi}{2}\right)\right)+3\bigg]\label{fidN}
\end{split}
\end{equation}

where $G+iF=\int\limits _{0}^{T}dt\sum_{j}r_{j}e^{in_{j}\xi t}$,    $A=-\int\limits_{0}^{T}dt g F$, and $\bar{n}$ is the mean occupation of the normal mode, assuming an initial thermal state. Expanding Eq. \eqref{fidN} in powers of $N$ we find that for a MS gate, the infidelities due to an inaccurate gate time, $T$, or normal-mode frequency, $\nu$, scale quadratically with $N$. Thus, using a gate that is robust to these type of errors is necessary for scaling up the ion-trap quantum processor. Similar scaling laws occur for $N$-qubit all-to-all entangling gates. Furthermore, the gate time typically scales as $\sqrt{N}$ resulting in an even increased sensitivity to errors, which can be mitigated only by coupling to additional normal-modes of motion of the ion-chain \cite{Shapira2020}.

By choosing $\{n_j,r_j\}$ correctly, one can use the additional degrees of freedom to suppress unwanted sources of infidelity, in an order-by-order manner. In our quantum information processor, we use two modulation terms rather than one which leads to a Cardioid-shaped trajectory of the normal-mode in phase space. A Cardioid gate eliminates the leading order contribution of carrier coupling and errors due to timing inaccuracies, and reduces errors due to normal-mode frequency inaccuracies and normal-mode heating.

While the most efficient implementation uses harmonics $n_j=1,2$, we realized the technique with the harmonics $n_j=2,3$ or $n_j=1,3$, in order to avoid errors caused by spurious frequencies due to third-order nonlinearities in the AOM or the RF amplifier. The double amplitude modulation is equivalent to a four-frequency field which is generated by driving an AOM with RF generated by a Keysight Trueform 33622a arbitrary waveform generator (AWG). 

In addition to the Cardioid modulation, we implement a Hahn echo pulse midway through the gate operation. For the MS gate, this implies a prolongation of the gate time by a factor of $\sqrt{2}$, but increases the gate robustness to $\sigma^z$ noise, which is our main source of decoherence. 

\begin{figure}
\centering
	\includegraphics[width=1\linewidth]{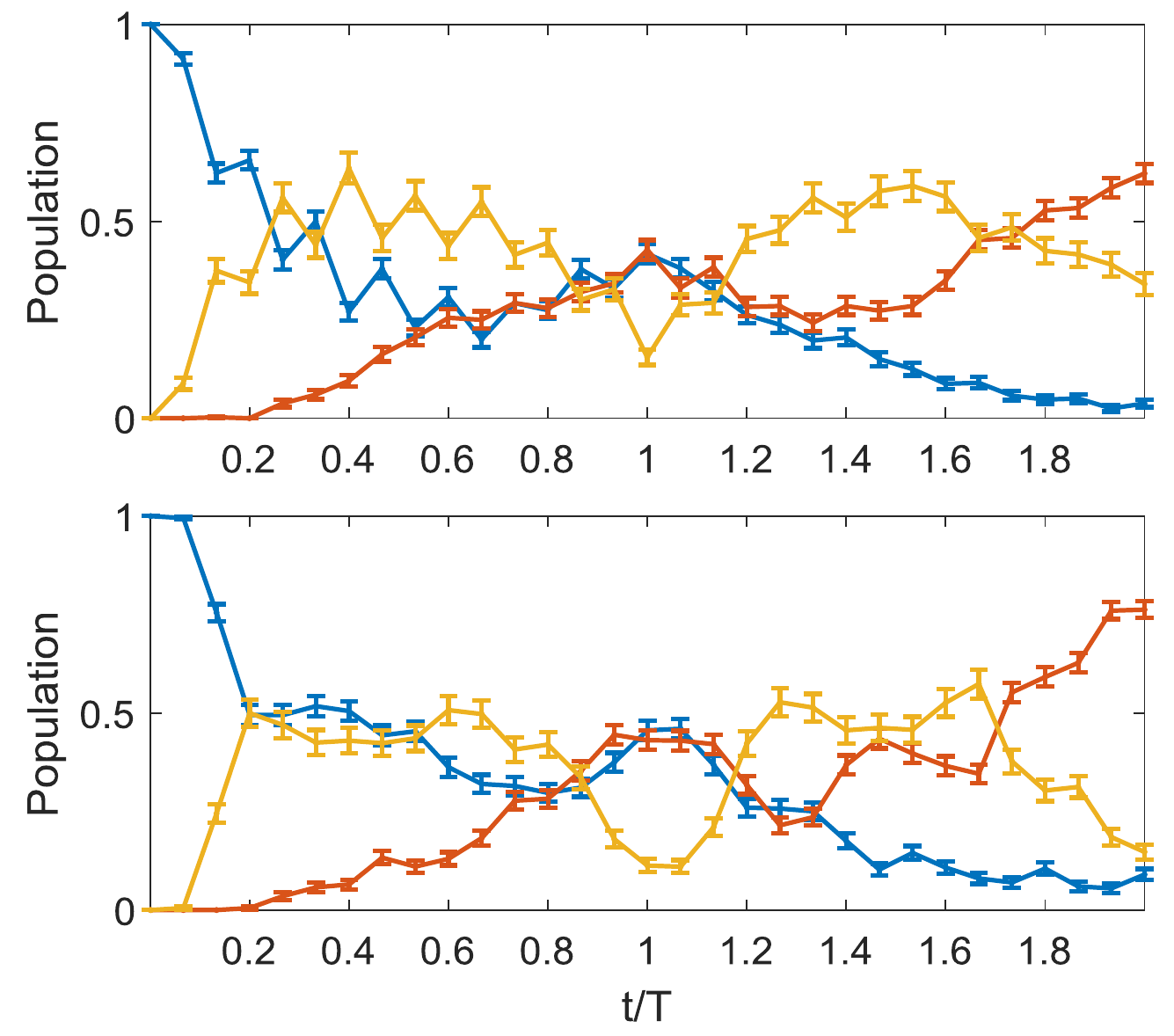}
	\caption[Dynamics of 4 ions all-to-all MS and Cardioid gates.]{Population of the ground state, $|1111\rangle$ (blue), and the fully excited state, $|0000\rangle$ (red) are shown. The system is initialized at $t=0$ to the ground state. Ideally, the population of these states at gate time, $t=T$, is $0.5$, implying a maximally entangled qubit state. In addition, the population of the 14 remaining states is shown (yellow), Ideally the population of these states at gate time is 0, indicating that the qubit state is disentangled from the motion. Error bars reflect shot noise due to 400 repetitions of each evolution time. In the MS gate (top), the population changes sharply at the gate time, indicating an increased sensitivity to pulse timing. While the Cardioid gate (bottom) approaches gate time smoothly, indicating a robust gate. Furthermore the Cardioid generates populations that are closer to the ideal case, due to reduced carrier coupling.}
	\label{dynamics4}
\end{figure} 

In order to characterize the fidelity of our entangling gates we implemented two-qubit Cardioid gates on all pairs of qubits in a 5-qubit chain. We encoded qubits on the $\ket{1}=5S_{\frac{1}{2},\frac{1}{2}}$ and $\ket{0}=4D_{\frac{5}{2},\frac{3}{2}}$ states, and used the $\ket{\tilde{1}}=S_{\frac{1}{2},-\frac{1}{2}}$ and $\ket{\tilde{0}}=4D_{\frac{5}{2},-\frac{3}{2}}$ states as a ``hidden subspace''; qubits encoded on the latter are spectrally decoupled from the entangling interaction. For each qubit pair, we first prepared the register in the $\ket{11111}$ state using optical pumping. We used individual addressing and global 674 nm pulses, as well as RF pulses on the Zeeman qubit, to hide all qubits but the targeted pair onto their corresponding hidden subspace, producing the state $\ket{\tilde{1}\tilde{1}\tilde{1}}\otimes\ket{11}$, where the latter two represent the target qubits. We then carried out a global Cardioid gate with an echo pulse,  ideally producing the maximally entangled Bell state $\ket{\psi_{Bell}}=\frac{1}{\sqrt{2}}(\ket{11}+i\ket{00})$ for these two qubits. We measured the state fidelity resulting from this procedure by measuring the populations and coherence of this two-qubit state through a parity scan.    


The results of our measurements are summarized in Fig. \ref{fidmatrix}. After correcting for the preparation cost of hiding three out of five ions, measured independently, as well as state measurement error, we reach an average fidelity of $0.973(2)$. The gate duration, including the $\sqrt{2}$ prolongation implied by the echo pulse addition, is approximately 200 $\mu s$.  These fidelities are comparable with those reported by state of the art leading trapped ion quantum computers \cite{figgatt2019parallel,pogorelov2021compact}. We believe that our major source of error is due to magnetic $\sigma^z$ noise, which is only partially mitigated by the echo pulse, as the optical qubit coherence time is limited to approximately two milliseconds, only a a single order of magnitude longer than the gate time. Hence, we expect significant improvements in fidelity with straightforward improvements to our experimental setup, such as the addition of a $\mu$-metal shield. 

Using the same methods described above, we also implemented a Cardioid all-to-all gate on a $N=4$ ion chain, and compared it with a conventional MS gate. Figure \ref{dynamics4} shows the dynamics of four ions under the MS (top) and Cardioid (bottom) all-to-all entanglement gates. Robustness of the Cardioid to pulse time errors is apparent as the evolution in the vicinity of $t=T$ is much smoother than that of the MS gate. Furthermore, the dynamics generated by the Cardioid gate is closer to the ideal case compared to MS.

\begin{figure}
\centering
	\includegraphics[width=1\linewidth]{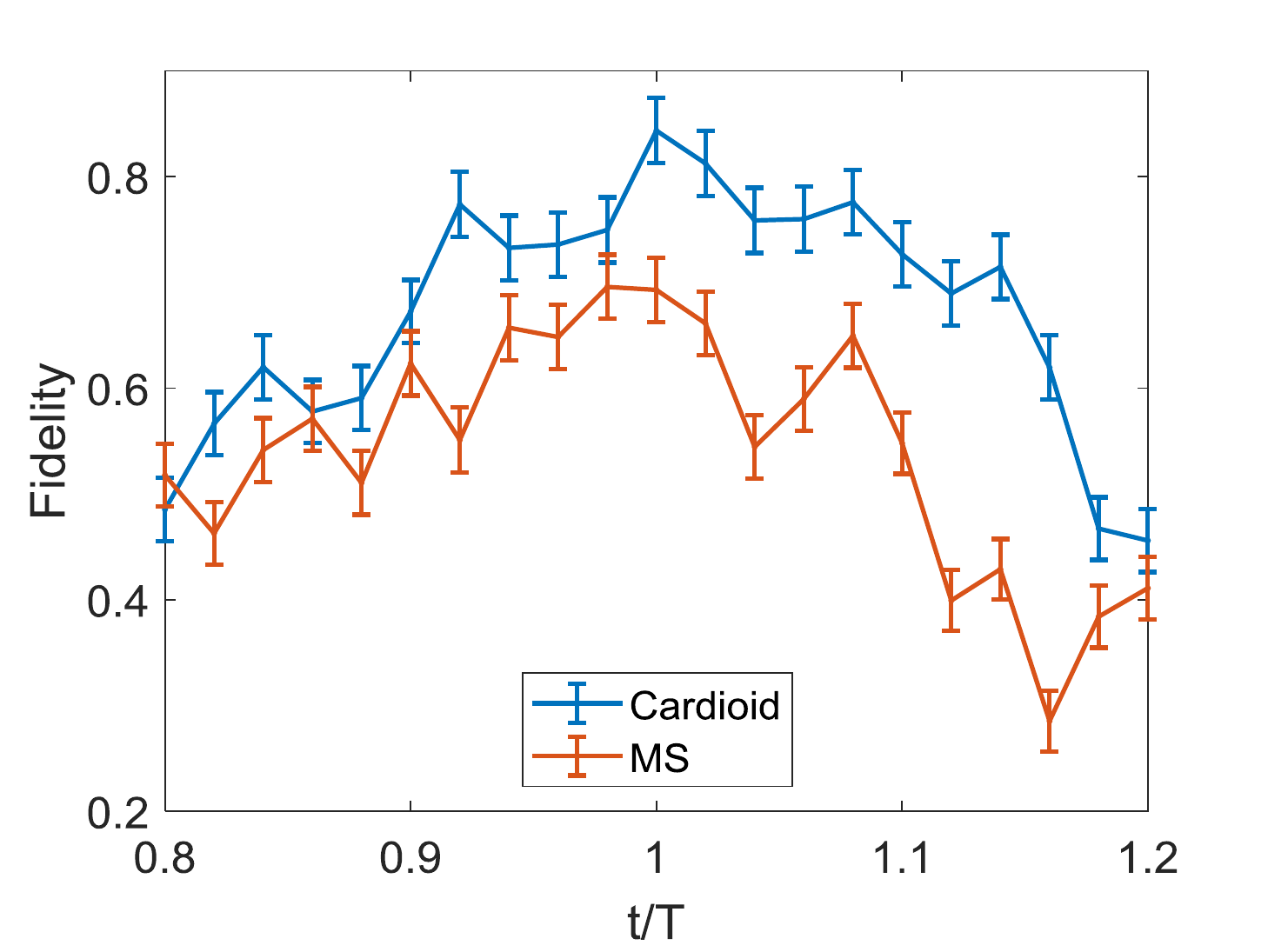}
	\caption[State fidelity for a 4-qubit GHZ state generated by MS and Cardioid gates for a 4 ions all-to-all entangling gate.]{The state fidelity for a 4-qubit GHZ state generated by the Cardioid (blue) and MS (red) gates is measured for different gate times. The Cardioid gate exhibits a higher peak fidelity and reduced sensitivity to pulse time errors compared to MS.}
	\label{fidrobust}
\end{figure}

To further demonstrate the robustness of the Cardioid to pulse time errors we measured the gate fidelity at different evolution times. Figure \ref{fidrobust} compares the 4-qubit gate fidelity of the Cardioid (blue) and MS (red) gates. Clearly the Cardioid exhibits a higher fidelity which is also less sensitive to incorrect gate times. The peak fidelties are $0.84(3)$ and $0.69(3)$ for Cardioid and MS gates respectively.

%% file: Sections/feedback.tex
Neither complete isolation of a quantum computer from the environment nor perfectly executed unitary operations are physically realizable goals, and errors will inevitably decohere the quantum computer states. Fortunately, a family of techniques known as quantum error correction (QEC) can enable fault-tolerant quantum computation for arbitrarily long computations given low enough error rates \cite{aharonov1997fault,knill1997theory}. QEC protects quantum information by encoding it non-locally, using highly entangled states. As the information is encoded over high order correlations, the otherwise harmful local noise does not destroy information, and the errors it inflicts can therefore be rectified. 

QEC requires measuring parts of the qubit register while maintaining the quantum coherence of the rest, and generating feedback operations conditioned on the measurement results. Generally, this is a technically difficult task: measurement is a highly dissipative process, strongly coupling parts of the quantum computer to the environment; at the same time, other (often neighboring) parts of the computer must remain completely decoupled from the environment. Furthermore, the measurement must be read out, and a conditional operation must be applied, all within the coherence time of the system. 

Standard state selective fluorescence detection on an EMCCD camera is ill-equipped to fulfill these demands.  Photon scattering from a laser beam overlapping the entire chain collapses the wavefunction of all qubits in the register. Even if the measurement beam is focused onto single ions, since photon collection is limited, typically thousands of photons have to be scattered before the state of the ion can be inferred with sufficient statistical certainty. Secondary photon scattering will lead to errors on other ions with high probability.

Furthermore, the readout of EMCCD cameras is typically slow since the transfer of charge across the camera pixel array takes time. Thus, in most camera-based readout, the processing of the detection results is done retroactively. Experiments are  almost always repeated many times, and in each iteration, the image acquired by the camera is saved onto a buffer. Only after all iterations are complete, the entire set of images is transferred from the buffer to a computer via USB, where image processing is performed to determine the measurement results. This retroactive processing is not adequate for real-time coherent feedback. Since the USB interface is slow and ultimately controlled by the CPU, simply reading out each measurement individually is not a viable solution to this problem. Furthermore, after readout, each image must be processed quickly, and the result should be fed back into the system within the coherence time of the qubits.


\begin{figure}
	\includegraphics[width=1\linewidth]{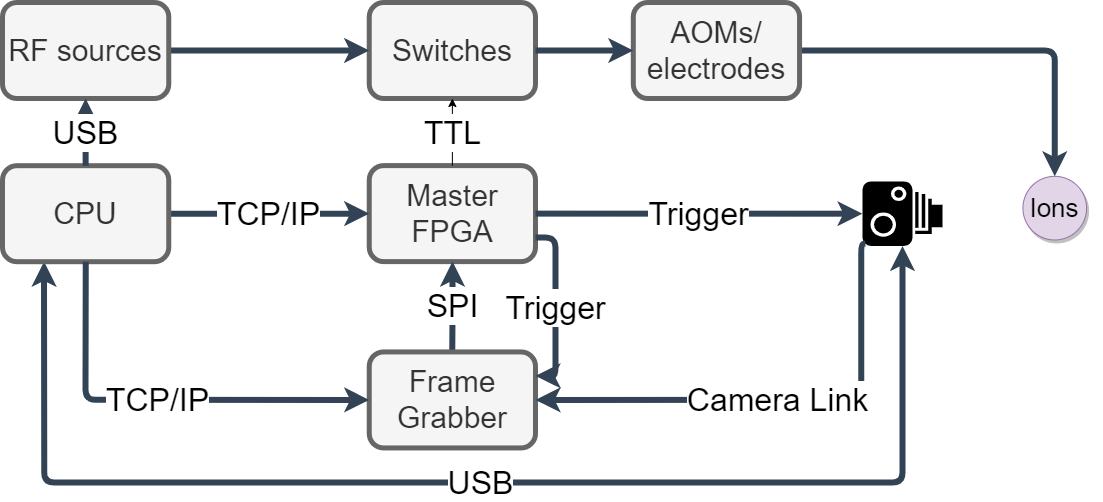}
	\caption[Quantum coherent feedback control scheme.]{The control scheme for the quantum coherent feedback module. The user controls the experiment using the CPU, which through USB determines parameters for the RF frequency sources and camera, and via TCP/IP uploads detection parameters to the frame grabber FPGA, and an experimental sequence onto the main FPGA. Once started, the main FPGA executes the sequence by controlling a set of TTL digital outputs, which can turn on or off electromagnetic fields addressing the ions, and can trigger camera detection. Live feedback is enabled by the addition of a frame grabber FPGA. The framegrabber receives real-time camera images through a Camera-Link communication protocol, processes the images according to parameters uploaded via TCP/IP, and outputs the results in SPI form to the main FPGA. The latter is programmed to continue the sequence according to these results.}
	\label{feedback_hardware}
\end{figure} 

\begin{figure*}
	\includegraphics[width=1\linewidth]{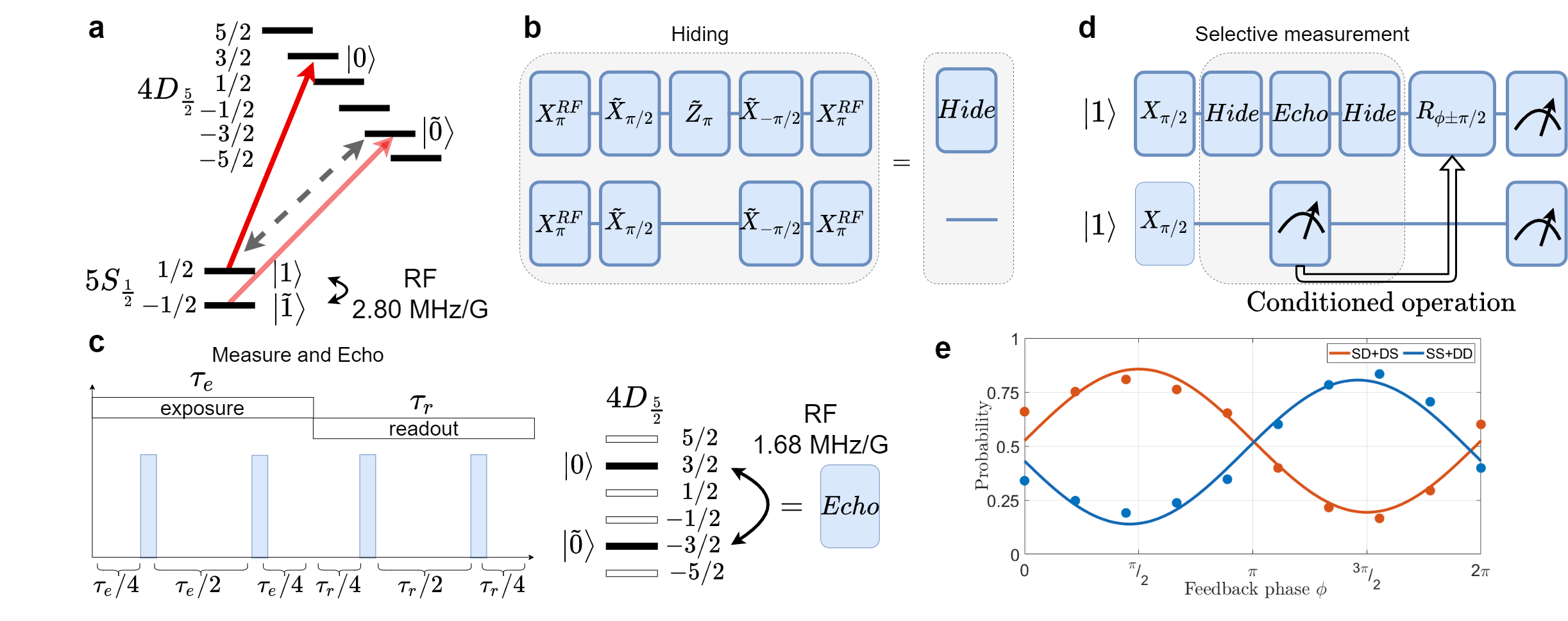}
	\caption[Measurement and coherence techniques for coherent feedback.]{(a) For qubits that should remain hidden from measurement, the $\ket{1}$ state is mapped onto $\ket{\tilde{0}}$; after the mapping procedure, these qubits are entirely encoded in the $4D_\frac{5}{2}$ manifold, and are decoupled from the measurement process. (b) Mapping is carried out by a series of operations utilizing optical control of the standard optical qubit $(\ket{0},\ket{1})$, a ``shadow'' optical qubit comprised of  $(\ket{\tilde{0}},\ket{\tilde{1}})$, and RF control of the $5S_\frac{1}{2}$ manifold qubit. Individual addressing pulses are used to single out the hidden ions. (c) After hiding, the remaining qubits are measured. During measurement the hidden qubits must maintain coherence in face of dephasing noise. We use RF pulses resonant with the $D$ manifold Zeeman splitting to flip between $\ket{0}$ and $\ket{\tilde{0}}$, echoing out the noise. We apply four $\pi$ pulses in a double CPMG configuration, such that noise during exposure and readout are decoupled separately. Typical times are $\tau_e\approx700 \mu s$ and $\tau_r\approx600 \mu s$. (d) In order to test coherent feedback, two qubits are initialized in the $\ket{1}$ state. We first apply a $\pi/2$ pulse on both qubits. Then, we apply a readout procedure, where one ion is hidden, the other is measured in parallel with the echo sequence shown in (c), and the hiding process is then reversed. Finally, a $\pi/2$ pulse is applied on the target qubit, with a phase that is controlled by the measurement result, and the two qubits are measured. (e) Parity measurement results are plotted as a function of $\phi$. For a non-conditioned pulse, or for total loss of coherence, we would expect a flat parity signal that is independent of phase. Instead, we measure an 0.83 success probability for the feedback operation, determined by the amplitude of the parity oscillations.}
	\label{feedback_fig}
\end{figure*}

In order to address these challenges, we developed a module and a procedure for live readout and real-time quantum-coherent feedback in our system. In order to measure only pre-selected qubits, we use individual addressing pulses to ``hide'' all other qubits by encoding them in levels which are completely decoupled from 422 nm scattering and therefore the measuring process. This is achieved by making use of another state in the six-dimensional Zeeman manifold of the $4D_\frac{5}{2}$ orbital. Specifically, while qubits to be measured remain encoded on the optical qubit states $\ket{1} = \ket{5S_{\frac{1}{2},\frac{1}{2}}}$ and $\ket{0} = \ket{4D_{\frac{5}{2},\frac{3}{2}}}$, hidden qubits are encoded into the $\ket{0} = \ket{4D_{\frac{5}{2},\frac{3}{2}}}$ and $\ket{\tilde{0}}=\ket{4D_{\frac{5}{2},-\frac{3}{2}}}$ states (see Fig. \ref{feedback_fig} (a)). The hiding process is realized using a series of RF and optical pulses, as shown in Fig. \ref{feedback_fig} (b). After the hiding protocol, we measure the remaining qubits by turning on the 422 and 1092 nm lasers, performing our standard measurement protocol, but reading out and processing the results in real time. Then, the hiding process is reversed, and an operation conditioned on the measurement result can be applied.

To implement real-time readout, processing and feedback, we realize a scheme which completely bypasses the camera buffer, the USB readout, the retroactive image processing, and in fact, any PC involvement. Fast readout is enabled by the Camera-Link port of the iXon Ultra camera. Camera-Link is a communication protocol that is designed for fast reading of images, and in the iXon Ultra outputs data before they reach the camera buffer. The acquired image is read out and processed in real-time by a dedicated framegrabber FPGA (National Instruments NI-PCIe-1473R), which comes with a built-in Camera-Link port. The framegrabber can perform image processing on the Camera-Link data on the fly, while information is still arriving. In this way, decisions on the outcome of the measurement, expressed as a bit state for each qubit, are made very quickly, and for some cases, qubit states are inferred before the complete image has even been read out by the analog-to-digital converter (ADC). The processing algorithm is pre-configured on the framegrabber FPGA in advance, and is identical to the algorithm realized in PC image processing: a set of pixels and a threshold value are allocated to each ion; for every image, the framegrabber sums over the values on pixels ascribed to each ion and determines if the result is larger than the corresponding threshold value; a binary string with the length of the ion chain is given as output.

In order for this fast determination of the ion state to be employed for coherent feedback, it must be available for use in real time by the main FPGA controlling all other parts of the experiment. Therefore, some communication between the two is necessary. We realized one-way synchronous (clock-controlled) serial communication using Serial Peripheral Interface (SPI). The framegrabber output is transferred to the main FPGA through SPI, while the main FPGA is configured to wait for the measurement results and to incorporate conditioned operations based on these results. In the opposite direction, the main FPGA sends a trigger to both the camera and the framegrabber at the beginning of every experiment, preparing them for an upcoming measurement \cite{Gazit:2020}. Our control scheme is shown diagrammatically in Fig. \ref{feedback_hardware}.

The speed in which the image is digitally read out by the camera, sent through Camera-Link to the framegrabber, processed, the results sent to the main FPGA, and the conditioned operation is triggered, is orders of magnitude shorter than either the acquisition or readout time (over 500 $\mu s$ each). Hence, this process introduces negligible time overhead.

The image acquisition time, including both EMCCD exposure and readout time, is typically longer than 1 ms and can be as high as 2 ms, during which require that the hidden ions maintain quantum coherence. However, the encoded states $(\ket{0},\ket{\tilde{0}})$ are highly susceptible to magnetic fields, and therefore magnetic noise in our lab limits the coherence times of the hidden qubits to about $500 \mu s$. This is strongly prohibitive for performing coherent feedback. In order to suppress noise, we apply dynamical decoupling (DD) pulses on the $4D_{\frac{5}{2}}$  manifold \emph{during} the measurement of the rest of the qubits. The DD pulses are realized by driving an RF magnetic field resonant with the $4D_{\frac{5}{2}}$ Zeeman splitting. This produces a $J_x$ Hamiltonian on the spin 5/2 Zeeman manifold, and for a correct time exactly realizes an effective $\pi$ pulse, translating the states $4D_{\frac{5}{2},m}\leftrightarrow 4D_{\frac{5}{2},-m}$. Thus,  this operation flips magnetic susceptibility and can be used as an echo pulse to suppress noise. Moreover, this driving has no bearing on the 422 nm measurement, as either optical qubit states remain within their subspace and appropriately maintain an identical measurement output. We realize the dynamical decoupling as a series of four echo pulses in a Carr-Purcell-Meiboom-Gill (CPMG) protocol \cite{carr1954effects,meiboom1958modified}.

\subsection{Demonstrating coherent feedback}

In order to verify and characterize the module's performance, we performed two tests for which real-time,  coherent, quantum feedback is necessary. Both experiments involve two qubits encoded on two ions: a control qubit, which is measured; and a target qubit, which is subject to a feedback operation. For both experiments, the operation on the target qubit is successful only if it is correctly conditioned on the result of the measured qubit, and if the target qubit remains coherent. For the first experiment the two qubits are in a separable state, while in the second experiment the ions are entangled prior to measurement.


In the first experiment, we prepared both qubits in the $\ket{1}$ state and applied a $\pi/2$ pulse, generating the separable state 
\begin{equation}
    \ket{\psi_1} = \frac{1}{2}\left( \ket{11}+i\ket{10}+i\ket{01}-\ket{00} \right).
\end{equation}
We then hid the target qubit, measured the control qubit, and reversed the hiding process, according to the protocol described earlier. Finally, another $\pi/2$ pulse was applied on the target qubit, with a phase that was conditioned on the measurement result: $\phi-\pi/2$ if $\ket{1}$ was measured, and $\phi+\pi/2$ if $\ket{0}$ was measured. As examples of the consequence of this conditioned operation, consider that in the case $\phi = \pi/2$, the target qubit will always be found in the opposite state to that of the control, whereas in the case of $\phi=-\pi/2$ the target and control qubit will be measured in the same state. Notice that if the operation is not conditioned on the measurement result, which is completely random, \emph{or} if it is conditioned, and the target qubit is no longer coherent, then the qubits will exhibit no correlations. More generally, we measured the correlation/anti-correlation of the control and target qubit measurement outcomes as a function of $\phi$. We successfully realized our desired correlation with a probability of 0.832(24), compared to random 0.5 probability, as shown in Fig. \ref{feedback_fig}e. The main source of error is likely the remainder magnetic field noise, along with imperfect implementation of DD RF pulses leading to population remaining in the $4D_{\frac{5}{2}}$ manifold.

\begin{figure}
\centering
	\includegraphics[width=1\linewidth]{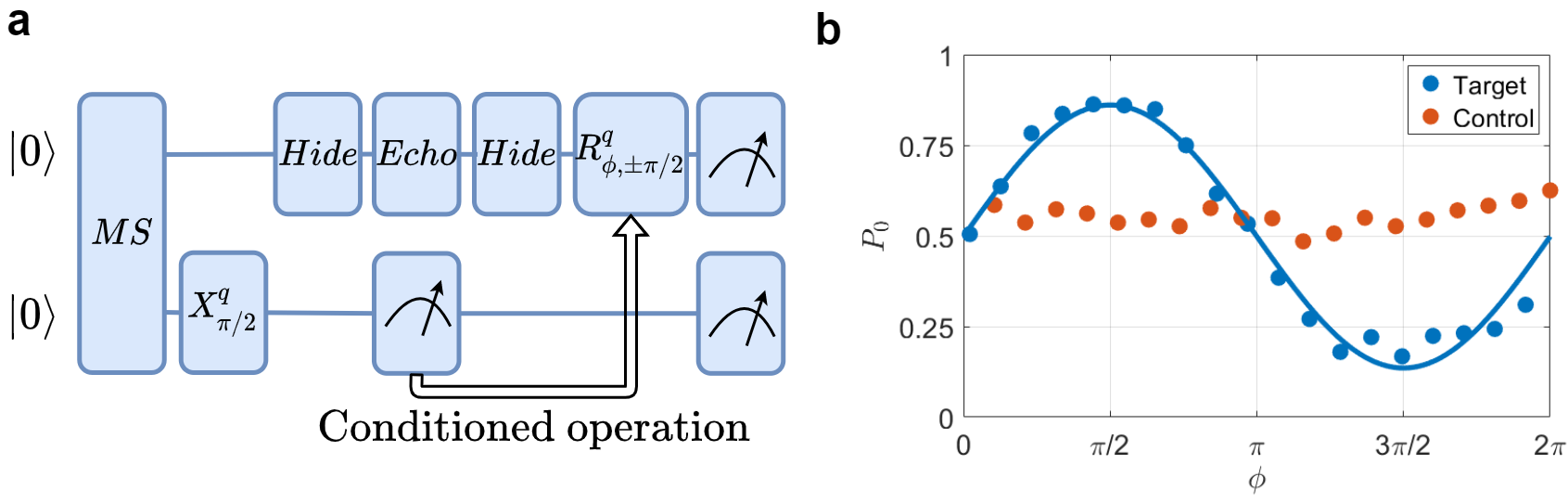}
	\caption[Coherent feedback on entangled state.]{ Coherent feedback on an entangled state. (a) A two qubit maximally entangled Bell state is prepared using an MS gate, followed by a $\pi/2$ pulse on the control qubit. Then, the target qubit is hidden, a measurement is performed along with echo pulses, and the target qubit is mapped back onto the optical qubit. A $\pi/2$ pulse with a phase $\phi-\pi/2$ or $\phi+\pi/2$ is applied conditioned on the measurement result. (b) Qubit population as a function of $\phi$. Due to the initial entanglement, the measurement collapses the target qubit onto a superposition state defined by the control qubit measurement result. This phase is effectively measured by applying the $\pi/2$ pulse. We measure an 0.862(18) success probability.}
	\label{feedback_ent}
\end{figure}

In a second experiment, we prepared the two qubits in the maximally entangled Bell state $\ket{\psi_{Bell}}=\frac{1}{\sqrt{2}}\left(\ket{11}+i\ket{00} \right)$ using the MS gate. This was followed by a $\pi/2$ pulse on the control qubit, leading to the state,
\begin{equation}
    \ket{\psi^\prime_{Bell}} =\frac{1}{\sqrt{2}}\left(\ket{\tilde{+}1}+i\ket{\tilde{-}0}\right),
\end{equation}
where $\ket{\tilde{\pm}}=\frac{1}{\sqrt{2}}(\ket{1}\pm i\ket{0})$ are Pauli $Y$ eigenstates. We then hid the target qubit, measured the control qubit, and reversed the hiding of the target qubit, with a sequence identical to that described above. 

The measurement projects the two-qubit state to either $\ket{1}\otimes\left(\frac{1}{\sqrt{2}}(\ket{1}+i\ket{0}) \right)=\ket{1\tilde{+}}$ or $\ket{0}\otimes\left(\frac{1}{\sqrt{2}}(\ket{1}-i\ket{0}) \right)=\ket{0\tilde{-}}$. The target qubit is therefore projected onto a superposition state with some well-defined phase, which is exactly opposite for the two cases. Finally, as in the previous experiment, we apply a $\pi/2$ pulse with phase $\phi-\pi/2$ if the control is measured as $\ket{1}$ and $\phi+\pi/2$ if the control is measured as $\ket{0}$. For conditioned coherent experiments, $\phi = \pi/2$ should always generate $\ket{0}$ on the target qubit, while $\phi = 3\pi/2$ should always generate $\ket{1}$. The measurement result is completely random, but because we exploit the correlation between the entangled qubits, the conditioned operation results in a deterministic outcome. Moreover, correct feedback relies on the phase coherence of the target qubit. For this measurement we report a success probability of 0.863(15), on par with the previous experiment, as can be seen in Fig.\ref{feedback_ent}.

While we demonstrated all necessary ingredient for performing quantum error correction with our system, our coherent feedback fidelities need to be improved. The main source of error is decoherence due to magnetic noise during the long readout time overhead $(\sim300 \mu s)$ of the EMCCD camera. Future systems, incorporating better-suited EMCCD cameras or PMT arrays and magnetic shielding and stabilization, will potentially increase fidelity. 

%% file: Sections/conclusion.tex
Trapped ion are among the leading candidate technologies for the realization of a large scale quantum computer. Such a large scale QC is likely to be composed out of an interconnected network of small quantum registers with high performance of universal quantum computing .

Here we have demonstrated such a small scale, five qubit, quantum register. In this register we have shown two important features. The first is the implementation of two-qubits gates that are robust against gate timing errors. Gates executed on pairs of ion-qubits out of a longer chain are more prone to errors than when executed on two-ion crystals. The use of robust gates is therefore important in order to reach high fidelity gates in longer crystals. The second feature is our demonstration of mid-circuit measurement and coherent feedback. The ability to perform mid-circuit measurement and feedback is at the heart of most quantum error-correction schemes and other protocols such as entanglement distillation or quantum teleprotation, all of which are inherent ingredients in large scale quantum computing.  

The improvement in performance of small scale quantum registers, as shown in this work, is an important step en-route the development of large scale quantum computing.

This work was performed with the support of the Israeli-Science Foundation. The Willner Family Leadership Institute for the Weizmann Institute of Science, The
Crown Photonics Center and the
Rosa and Emilio Segre Research Award.

%% file: main.bbl
\begin{thebibliography}{45}%
\makeatletter
\providecommand \@ifxundefined [1]{%
 \@ifx{#1\undefined}
}%
\providecommand \@ifnum [1]{%
 \ifnum #1\expandafter \@firstoftwo
 \else \expandafter \@secondoftwo
 \fi
}%
\providecommand \@ifx [1]{%
 \ifx #1\expandafter \@firstoftwo
 \else \expandafter \@secondoftwo
 \fi
}%
\providecommand \natexlab [1]{#1}%
\providecommand \enquote  [1]{``#1''}%
\providecommand \bibnamefont  [1]{#1}%
\providecommand \bibfnamefont [1]{#1}%
\providecommand \citenamefont [1]{#1}%
\providecommand \href@noop [0]{\@secondoftwo}%
\providecommand \href [0]{\begingroup \@sanitize@url \@href}%
\providecommand \@href[1]{\@@startlink{#1}\@@href}%
\providecommand \@@href[1]{\endgroup#1\@@endlink}%
\providecommand \@sanitize@url [0]{\catcode `\\12\catcode `\$12\catcode
  `\&12\catcode `\#12\catcode `\^12\catcode `\_12\catcode `\%12\relax}%
\providecommand \@@startlink[1]{}%
\providecommand \@@endlink[0]{}%
\providecommand \url  [0]{\begingroup\@sanitize@url \@url }%
\providecommand \@url [1]{\endgroup\@href {#1}{\urlprefix }}%
\providecommand \urlprefix  [0]{URL }%
\providecommand \Eprint [0]{\href }%
\providecommand \doibase [0]{http://dx.doi.org/}%
\providecommand \selectlanguage [0]{\@gobble}%
\providecommand \bibinfo  [0]{\@secondoftwo}%
\providecommand \bibfield  [0]{\@secondoftwo}%
\providecommand \translation [1]{[#1]}%
\providecommand \BibitemOpen [0]{}%
\providecommand \bibitemStop [0]{}%
\providecommand \bibitemNoStop [0]{.\EOS\space}%
\providecommand \EOS [0]{\spacefactor3000\relax}%
\providecommand \BibitemShut  [1]{\csname bibitem#1\endcsname}%
\let\auto@bib@innerbib\@empty
\bibitem [{\citenamefont {Ozeri}(2011)}]{Ozeri2011Toolbox}%
  \BibitemOpen
  \bibfield  {author} {\bibinfo {author} {\bibfnamefont {R.}~\bibnamefont
  {Ozeri}},\ }\href {\doibase 0.1080/00107514.2011.603578} {\bibfield
  {journal} {\bibinfo  {journal} {Contemporary Physics}\ }\textbf {\bibinfo
  {volume} {52}},\ \bibinfo {pages} {531} (\bibinfo {year} {2011})}\BibitemShut
  {NoStop}%
\bibitem [{\citenamefont {Kielpinski}\ \emph {et~al.}(2002)\citenamefont
  {Kielpinski}, \citenamefont {Monroe},\ and\ \citenamefont
  {Wineland}}]{Kielpinski2002}%
  \BibitemOpen
  \bibfield  {author} {\bibinfo {author} {\bibfnamefont {D.}~\bibnamefont
  {Kielpinski}}, \bibinfo {author} {\bibfnamefont {C.}~\bibnamefont {Monroe}},
  \ and\ \bibinfo {author} {\bibfnamefont {D.~J.}\ \bibnamefont {Wineland}},\
  }\href {\doibase 10.1038/nature00784} {\bibfield  {journal} {\bibinfo
  {journal} {Nature}\ }\textbf {\bibinfo {volume} {417}},\ \bibinfo {pages}
  {709} (\bibinfo {year} {2002})}\BibitemShut {NoStop}%
\bibitem [{\citenamefont {Wan}\ \emph {et~al.}(2019)\citenamefont {Wan},
  \citenamefont {Kienzler}, \citenamefont {Erickson}, \citenamefont {Mayer},
  \citenamefont {Tan}, \citenamefont {Wu}, \citenamefont {Vasconcelos},
  \citenamefont {Glancy}, \citenamefont {Knill}, \citenamefont {Wineland},
  \citenamefont {Wilson},\ and\ \citenamefont {Leibfried}}]{Wan2019}%
  \BibitemOpen
  \bibfield  {author} {\bibinfo {author} {\bibfnamefont {Y.}~\bibnamefont
  {Wan}}, \bibinfo {author} {\bibfnamefont {D.}~\bibnamefont {Kienzler}},
  \bibinfo {author} {\bibfnamefont {S.~D.}\ \bibnamefont {Erickson}}, \bibinfo
  {author} {\bibfnamefont {K.~H.}\ \bibnamefont {Mayer}}, \bibinfo {author}
  {\bibfnamefont {T.~R.}\ \bibnamefont {Tan}}, \bibinfo {author} {\bibfnamefont
  {J.~J.}\ \bibnamefont {Wu}}, \bibinfo {author} {\bibfnamefont {H.~M.}\
  \bibnamefont {Vasconcelos}}, \bibinfo {author} {\bibfnamefont
  {S.}~\bibnamefont {Glancy}}, \bibinfo {author} {\bibfnamefont
  {E.}~\bibnamefont {Knill}}, \bibinfo {author} {\bibfnamefont {D.~J.}\
  \bibnamefont {Wineland}}, \bibinfo {author} {\bibfnamefont {A.~C.}\
  \bibnamefont {Wilson}}, \ and\ \bibinfo {author} {\bibfnamefont
  {D.}~\bibnamefont {Leibfried}},\ }\href {\doibase 10.1126/science.aaw9415}
  {\bibfield  {journal} {\bibinfo  {journal} {Science}\ }\textbf {\bibinfo
  {volume} {364}},\ \bibinfo {pages} {875} (\bibinfo {year}
  {2019})}\BibitemShut {NoStop}%
\bibitem [{\citenamefont {Pino}\ \emph
  {et~al.}(2021{\natexlab{a}})\citenamefont {Pino}, \citenamefont {Dreiling},
  \citenamefont {Figgatt}, \citenamefont {Gaebler}, \citenamefont {Moses},
  \citenamefont {Allman}, \citenamefont {Baldwin}, \citenamefont {Foss-Feig},
  \citenamefont {Hayes}, \citenamefont {Mayer},\ and\ \citenamefont
  {et~al.}}]{Pino_2021}%
  \BibitemOpen
  \bibfield  {author} {\bibinfo {author} {\bibfnamefont {J.~M.}\ \bibnamefont
  {Pino}}, \bibinfo {author} {\bibfnamefont {J.~M.}\ \bibnamefont {Dreiling}},
  \bibinfo {author} {\bibfnamefont {C.}~\bibnamefont {Figgatt}}, \bibinfo
  {author} {\bibfnamefont {J.~P.}\ \bibnamefont {Gaebler}}, \bibinfo {author}
  {\bibfnamefont {S.~A.}\ \bibnamefont {Moses}}, \bibinfo {author}
  {\bibfnamefont {M.~S.}\ \bibnamefont {Allman}}, \bibinfo {author}
  {\bibfnamefont {C.~H.}\ \bibnamefont {Baldwin}}, \bibinfo {author}
  {\bibfnamefont {M.}~\bibnamefont {Foss-Feig}}, \bibinfo {author}
  {\bibfnamefont {D.}~\bibnamefont {Hayes}}, \bibinfo {author} {\bibfnamefont
  {K.}~\bibnamefont {Mayer}}, \ and\ \bibinfo {author} {\bibnamefont
  {et~al.}},\ }\href {\doibase 10.1038/s41586-021-03318-4} {\bibfield
  {journal} {\bibinfo  {journal} {Nature}\ }\textbf {\bibinfo {volume} {592}},\
  \bibinfo {pages} {209–213} (\bibinfo {year}
  {2021}{\natexlab{a}})}\BibitemShut {NoStop}%
\bibitem [{\citenamefont {Monroe}\ \emph {et~al.}(2014)\citenamefont {Monroe},
  \citenamefont {Raussendorf}, \citenamefont {Ruthven}, \citenamefont {Brown},
  \citenamefont {Maunz}, \citenamefont {Duan},\ and\ \citenamefont
  {Kim}}]{monroe2014large}%
  \BibitemOpen
  \bibfield  {author} {\bibinfo {author} {\bibfnamefont {C.}~\bibnamefont
  {Monroe}}, \bibinfo {author} {\bibfnamefont {R.}~\bibnamefont {Raussendorf}},
  \bibinfo {author} {\bibfnamefont {A.}~\bibnamefont {Ruthven}}, \bibinfo
  {author} {\bibfnamefont {K.~R.}\ \bibnamefont {Brown}}, \bibinfo {author}
  {\bibfnamefont {P.}~\bibnamefont {Maunz}}, \bibinfo {author} {\bibfnamefont
  {L.-M.}\ \bibnamefont {Duan}}, \ and\ \bibinfo {author} {\bibfnamefont
  {J.}~\bibnamefont {Kim}},\ }\href {\doibase 10.1103/PhysRevA.89.022317}
  {\bibfield  {journal} {\bibinfo  {journal} {Phys. Rev. A}\ }\textbf {\bibinfo
  {volume} {89}},\ \bibinfo {pages} {022317} (\bibinfo {year}
  {2014})}\BibitemShut {NoStop}%
\bibitem [{\citenamefont {Hucul}\ \emph {et~al.}(2015)\citenamefont {Hucul},
  \citenamefont {Inlek}, \citenamefont {Vittorini}, \citenamefont {Crocker},
  \citenamefont {Debnath}, \citenamefont {Clark},\ and\ \citenamefont
  {Monroe}}]{hucul2015modular}%
  \BibitemOpen
  \bibfield  {author} {\bibinfo {author} {\bibfnamefont {D.}~\bibnamefont
  {Hucul}}, \bibinfo {author} {\bibfnamefont {I.~V.}\ \bibnamefont {Inlek}},
  \bibinfo {author} {\bibfnamefont {G.}~\bibnamefont {Vittorini}}, \bibinfo
  {author} {\bibfnamefont {C.}~\bibnamefont {Crocker}}, \bibinfo {author}
  {\bibfnamefont {S.}~\bibnamefont {Debnath}}, \bibinfo {author} {\bibfnamefont
  {S.~M.}\ \bibnamefont {Clark}}, \ and\ \bibinfo {author} {\bibfnamefont
  {C.}~\bibnamefont {Monroe}},\ }\href@noop {} {\bibfield  {journal} {\bibinfo
  {journal} {Nature Physics}\ }\textbf {\bibinfo {volume} {11}},\ \bibinfo
  {pages} {37} (\bibinfo {year} {2015})}\BibitemShut {NoStop}%
\bibitem [{\citenamefont {Linke}\ \emph {et~al.}(2017)\citenamefont {Linke},
  \citenamefont {Maslov}, \citenamefont {Roetteler}, \citenamefont {Debnath},
  \citenamefont {Figgatt}, \citenamefont {Landsman}, \citenamefont {Wright},\
  and\ \citenamefont {Monroe}}]{Linke3305}%
  \BibitemOpen
  \bibfield  {author} {\bibinfo {author} {\bibfnamefont {N.~M.}\ \bibnamefont
  {Linke}}, \bibinfo {author} {\bibfnamefont {D.}~\bibnamefont {Maslov}},
  \bibinfo {author} {\bibfnamefont {M.}~\bibnamefont {Roetteler}}, \bibinfo
  {author} {\bibfnamefont {S.}~\bibnamefont {Debnath}}, \bibinfo {author}
  {\bibfnamefont {C.}~\bibnamefont {Figgatt}}, \bibinfo {author} {\bibfnamefont
  {K.~A.}\ \bibnamefont {Landsman}}, \bibinfo {author} {\bibfnamefont
  {K.}~\bibnamefont {Wright}}, \ and\ \bibinfo {author} {\bibfnamefont
  {C.}~\bibnamefont {Monroe}},\ }\href {\doibase 10.1073/pnas.1618020114}
  {\bibfield  {journal} {\bibinfo  {journal} {Proceedings of the National
  Academy of Sciences}\ }\textbf {\bibinfo {volume} {114}},\ \bibinfo {pages}
  {3305} (\bibinfo {year} {2017})},\ \Eprint
  {http://arxiv.org/abs/https://www.pnas.org/content/114/13/3305.full.pdf}
  {https://www.pnas.org/content/114/13/3305.full.pdf} \BibitemShut {NoStop}%
\bibitem [{\citenamefont {Egan}\ \emph {et~al.}(2020)\citenamefont {Egan},
  \citenamefont {Debroy}, \citenamefont {Noel}, \citenamefont {Risinger},
  \citenamefont {Zhu}, \citenamefont {Biswas}, \citenamefont {Newman},
  \citenamefont {Li}, \citenamefont {Brown}, \citenamefont {Cetina} \emph
  {et~al.}}]{egan2020fault}%
  \BibitemOpen
  \bibfield  {author} {\bibinfo {author} {\bibfnamefont {L.}~\bibnamefont
  {Egan}}, \bibinfo {author} {\bibfnamefont {D.~M.}\ \bibnamefont {Debroy}},
  \bibinfo {author} {\bibfnamefont {C.}~\bibnamefont {Noel}}, \bibinfo {author}
  {\bibfnamefont {A.}~\bibnamefont {Risinger}}, \bibinfo {author}
  {\bibfnamefont {D.}~\bibnamefont {Zhu}}, \bibinfo {author} {\bibfnamefont
  {D.}~\bibnamefont {Biswas}}, \bibinfo {author} {\bibfnamefont
  {M.}~\bibnamefont {Newman}}, \bibinfo {author} {\bibfnamefont
  {M.}~\bibnamefont {Li}}, \bibinfo {author} {\bibfnamefont {K.~R.}\
  \bibnamefont {Brown}}, \bibinfo {author} {\bibfnamefont {M.}~\bibnamefont
  {Cetina}},  \emph {et~al.},\ }\href@noop {} {\bibfield  {journal} {\bibinfo
  {journal} {arXiv preprint arXiv:2009.11482}\ } (\bibinfo {year}
  {2020})}\BibitemShut {NoStop}%
\bibitem [{\citenamefont {Schindler}\ \emph {et~al.}(2013)\citenamefont
  {Schindler}, \citenamefont {Nigg}, \citenamefont {Monz}, \citenamefont
  {Barreiro}, \citenamefont {Martinez}, \citenamefont {Wang}, \citenamefont
  {Quint}, \citenamefont {Brandl}, \citenamefont {Nebendahl}, \citenamefont
  {Roos} \emph {et~al.}}]{Schindler2013}%
  \BibitemOpen
  \bibfield  {author} {\bibinfo {author} {\bibfnamefont {P.}~\bibnamefont
  {Schindler}}, \bibinfo {author} {\bibfnamefont {D.}~\bibnamefont {Nigg}},
  \bibinfo {author} {\bibfnamefont {T.}~\bibnamefont {Monz}}, \bibinfo {author}
  {\bibfnamefont {J.~T.}\ \bibnamefont {Barreiro}}, \bibinfo {author}
  {\bibfnamefont {E.}~\bibnamefont {Martinez}}, \bibinfo {author}
  {\bibfnamefont {S.~X.}\ \bibnamefont {Wang}}, \bibinfo {author}
  {\bibfnamefont {S.}~\bibnamefont {Quint}}, \bibinfo {author} {\bibfnamefont
  {M.~F.}\ \bibnamefont {Brandl}}, \bibinfo {author} {\bibfnamefont
  {V.}~\bibnamefont {Nebendahl}}, \bibinfo {author} {\bibfnamefont {C.~F.}\
  \bibnamefont {Roos}},  \emph {et~al.},\ }\href@noop {} {\bibfield  {journal}
  {\bibinfo  {journal} {New Journal of Physics}\ }\textbf {\bibinfo {volume}
  {15}},\ \bibinfo {pages} {123012} (\bibinfo {year} {2013})}\BibitemShut
  {NoStop}%
\bibitem [{\citenamefont {Pogorelov}\ \emph
  {et~al.}(2021{\natexlab{a}})\citenamefont {Pogorelov}, \citenamefont
  {Feldker}, \citenamefont {Marciniak}, \citenamefont {Postler}, \citenamefont
  {Jacob}, \citenamefont {Krieglsteiner}, \citenamefont {Podlesnic},
  \citenamefont {Meth}, \citenamefont {Negnevitsky}, \citenamefont {Stadler},
  \citenamefont {H\"ofer}, \citenamefont {W\"achter}, \citenamefont
  {Lakhmanskiy}, \citenamefont {Blatt}, \citenamefont {Schindler},\ and\
  \citenamefont {Monz}}]{IQT2021}%
  \BibitemOpen
  \bibfield  {author} {\bibinfo {author} {\bibfnamefont {I.}~\bibnamefont
  {Pogorelov}}, \bibinfo {author} {\bibfnamefont {T.}~\bibnamefont {Feldker}},
  \bibinfo {author} {\bibfnamefont {C.~D.}\ \bibnamefont {Marciniak}}, \bibinfo
  {author} {\bibfnamefont {L.}~\bibnamefont {Postler}}, \bibinfo {author}
  {\bibfnamefont {G.}~\bibnamefont {Jacob}}, \bibinfo {author} {\bibfnamefont
  {O.}~\bibnamefont {Krieglsteiner}}, \bibinfo {author} {\bibfnamefont
  {V.}~\bibnamefont {Podlesnic}}, \bibinfo {author} {\bibfnamefont
  {M.}~\bibnamefont {Meth}}, \bibinfo {author} {\bibfnamefont {V.}~\bibnamefont
  {Negnevitsky}}, \bibinfo {author} {\bibfnamefont {M.}~\bibnamefont
  {Stadler}}, \bibinfo {author} {\bibfnamefont {B.}~\bibnamefont {H\"ofer}},
  \bibinfo {author} {\bibfnamefont {C.}~\bibnamefont {W\"achter}}, \bibinfo
  {author} {\bibfnamefont {K.}~\bibnamefont {Lakhmanskiy}}, \bibinfo {author}
  {\bibfnamefont {R.}~\bibnamefont {Blatt}}, \bibinfo {author} {\bibfnamefont
  {P.}~\bibnamefont {Schindler}}, \ and\ \bibinfo {author} {\bibfnamefont
  {T.}~\bibnamefont {Monz}},\ }\href {\doibase 10.1103/PRXQuantum.2.020343}
  {\bibfield  {journal} {\bibinfo  {journal} {PRX Quantum}\ }\textbf {\bibinfo
  {volume} {2}},\ \bibinfo {pages} {020343} (\bibinfo {year}
  {2021}{\natexlab{a}})}\BibitemShut {NoStop}%
\bibitem [{\citenamefont {Debnath}\ \emph {et~al.}(2016)\citenamefont
  {Debnath}, \citenamefont {Linke}, \citenamefont {Figgatt}, \citenamefont
  {Landsman}, \citenamefont {Wright},\ and\ \citenamefont
  {Monroe}}]{Debnath2016}%
  \BibitemOpen
  \bibfield  {author} {\bibinfo {author} {\bibfnamefont {S.}~\bibnamefont
  {Debnath}}, \bibinfo {author} {\bibfnamefont {N.~M.}\ \bibnamefont {Linke}},
  \bibinfo {author} {\bibfnamefont {C.}~\bibnamefont {Figgatt}}, \bibinfo
  {author} {\bibfnamefont {K.~A.}\ \bibnamefont {Landsman}}, \bibinfo {author}
  {\bibfnamefont {K.}~\bibnamefont {Wright}}, \ and\ \bibinfo {author}
  {\bibfnamefont {C.}~\bibnamefont {Monroe}},\ }\href {\doibase
  10.1038/nature18648} {\bibfield  {journal} {\bibinfo  {journal} {Nature}\
  }\textbf {\bibinfo {volume} {536}},\ \bibinfo {pages} {63} (\bibinfo {year}
  {2016})}\BibitemShut {NoStop}%
\bibitem [{\citenamefont {Wright}\ \emph {et~al.}(2019)\citenamefont {Wright},
  \citenamefont {Beck}, \citenamefont {Debnath}, \citenamefont {Amini},
  \citenamefont {Nam}, \citenamefont {Grzesiak}, \citenamefont {Chen},
  \citenamefont {Pisenti}, \citenamefont {Chmielewski}, \citenamefont
  {Collins}, \citenamefont {Hudek}, \citenamefont {Mizrahi}, \citenamefont
  {Wong-Campos}, \citenamefont {Allen}, \citenamefont {Apisdorf}, \citenamefont
  {Solomon}, \citenamefont {Williams}, \citenamefont {Ducore}, \citenamefont
  {Blinov}, \citenamefont {Kreikemeier}, \citenamefont {Chaplin}, \citenamefont
  {Keesan}, \citenamefont {Monroe},\ and\ \citenamefont {Kim}}]{IonQ2019}%
  \BibitemOpen
  \bibfield  {author} {\bibinfo {author} {\bibfnamefont {K.}~\bibnamefont
  {Wright}}, \bibinfo {author} {\bibfnamefont {K.~M.}\ \bibnamefont {Beck}},
  \bibinfo {author} {\bibfnamefont {S.}~\bibnamefont {Debnath}}, \bibinfo
  {author} {\bibfnamefont {J.~M.}\ \bibnamefont {Amini}}, \bibinfo {author}
  {\bibfnamefont {Y.}~\bibnamefont {Nam}}, \bibinfo {author} {\bibfnamefont
  {N.}~\bibnamefont {Grzesiak}}, \bibinfo {author} {\bibfnamefont {J.-S.}\
  \bibnamefont {Chen}}, \bibinfo {author} {\bibfnamefont {N.~C.}\ \bibnamefont
  {Pisenti}}, \bibinfo {author} {\bibfnamefont {M.}~\bibnamefont
  {Chmielewski}}, \bibinfo {author} {\bibfnamefont {C.}~\bibnamefont
  {Collins}}, \bibinfo {author} {\bibfnamefont {K.~M.}\ \bibnamefont {Hudek}},
  \bibinfo {author} {\bibfnamefont {J.}~\bibnamefont {Mizrahi}}, \bibinfo
  {author} {\bibfnamefont {J.~D.}\ \bibnamefont {Wong-Campos}}, \bibinfo
  {author} {\bibfnamefont {S.}~\bibnamefont {Allen}}, \bibinfo {author}
  {\bibfnamefont {J.}~\bibnamefont {Apisdorf}}, \bibinfo {author}
  {\bibfnamefont {P.}~\bibnamefont {Solomon}}, \bibinfo {author} {\bibfnamefont
  {M.}~\bibnamefont {Williams}}, \bibinfo {author} {\bibfnamefont {A.~M.}\
  \bibnamefont {Ducore}}, \bibinfo {author} {\bibfnamefont {A.}~\bibnamefont
  {Blinov}}, \bibinfo {author} {\bibfnamefont {S.~M.}\ \bibnamefont
  {Kreikemeier}}, \bibinfo {author} {\bibfnamefont {V.}~\bibnamefont
  {Chaplin}}, \bibinfo {author} {\bibfnamefont {M.}~\bibnamefont {Keesan}},
  \bibinfo {author} {\bibfnamefont {C.}~\bibnamefont {Monroe}}, \ and\ \bibinfo
  {author} {\bibfnamefont {J.}~\bibnamefont {Kim}},\ }\href {\doibase
  10.1038/s41467-019-13534-2} {\bibfield  {journal} {\bibinfo  {journal}
  {Nature Communications}\ }\textbf {\bibinfo {volume} {10}},\ \bibinfo {pages}
  {5464} (\bibinfo {year} {2019})}\BibitemShut {NoStop}%
\bibitem [{\citenamefont {Lu}\ \emph {et~al.}(2019)\citenamefont {Lu},
  \citenamefont {Zhang}, \citenamefont {Zhang}, \citenamefont {Chen},
  \citenamefont {Shen}, \citenamefont {Zhang}, \citenamefont {Zhang},\ and\
  \citenamefont {Kim}}]{Tsinghua2019}%
  \BibitemOpen
  \bibfield  {author} {\bibinfo {author} {\bibfnamefont {Y.}~\bibnamefont
  {Lu}}, \bibinfo {author} {\bibfnamefont {S.}~\bibnamefont {Zhang}}, \bibinfo
  {author} {\bibfnamefont {K.}~\bibnamefont {Zhang}}, \bibinfo {author}
  {\bibfnamefont {W.}~\bibnamefont {Chen}}, \bibinfo {author} {\bibfnamefont
  {Y.}~\bibnamefont {Shen}}, \bibinfo {author} {\bibfnamefont {J.}~\bibnamefont
  {Zhang}}, \bibinfo {author} {\bibfnamefont {J.-N.}\ \bibnamefont {Zhang}}, \
  and\ \bibinfo {author} {\bibfnamefont {K.}~\bibnamefont {Kim}},\ }\href
  {\doibase 10.1038/s41586-019-1428-4} {\bibfield  {journal} {\bibinfo
  {journal} {Nature}\ }\textbf {\bibinfo {volume} {572}},\ \bibinfo {pages}
  {363} (\bibinfo {year} {2019})}\BibitemShut {NoStop}%
\bibitem [{\citenamefont {Home}\ \emph {et~al.}(2009)\citenamefont {Home},
  \citenamefont {Hanneke}, \citenamefont {Jost}, \citenamefont {Amini},
  \citenamefont {Leibfried},\ and\ \citenamefont {Wineland}}]{NIST2009}%
  \BibitemOpen
  \bibfield  {author} {\bibinfo {author} {\bibfnamefont {J.~P.}\ \bibnamefont
  {Home}}, \bibinfo {author} {\bibfnamefont {D.}~\bibnamefont {Hanneke}},
  \bibinfo {author} {\bibfnamefont {J.~D.}\ \bibnamefont {Jost}}, \bibinfo
  {author} {\bibfnamefont {J.~M.}\ \bibnamefont {Amini}}, \bibinfo {author}
  {\bibfnamefont {D.}~\bibnamefont {Leibfried}}, \ and\ \bibinfo {author}
  {\bibfnamefont {D.~J.}\ \bibnamefont {Wineland}},\ }\href {\doibase
  10.1126/science.1177077} {\bibfield  {journal} {\bibinfo  {journal}
  {Science}\ }\textbf {\bibinfo {volume} {325}},\ \bibinfo {pages} {1227}
  (\bibinfo {year} {2009})},\ \Eprint
  {http://arxiv.org/abs/https://www.science.org/doi/pdf/10.1126/science.1177077}
  {https://www.science.org/doi/pdf/10.1126/science.1177077} \BibitemShut
  {NoStop}%
\bibitem [{\citenamefont {Negnevitsky}\ \emph {et~al.}(2018)\citenamefont
  {Negnevitsky}, \citenamefont {Marinelli}, \citenamefont {Mehta},
  \citenamefont {Lo}, \citenamefont {Flühmann},\ and\ \citenamefont
  {Home}}]{ETH2018}%
  \BibitemOpen
  \bibfield  {author} {\bibinfo {author} {\bibfnamefont {V.}~\bibnamefont
  {Negnevitsky}}, \bibinfo {author} {\bibfnamefont {M.}~\bibnamefont
  {Marinelli}}, \bibinfo {author} {\bibfnamefont {K.~K.}\ \bibnamefont
  {Mehta}}, \bibinfo {author} {\bibfnamefont {H.-Y.}\ \bibnamefont {Lo}},
  \bibinfo {author} {\bibfnamefont {C.}~\bibnamefont {Flühmann}}, \ and\
  \bibinfo {author} {\bibfnamefont {J.~P.}\ \bibnamefont {Home}},\ }\href
  {\doibase 10.1038/s41586-018-0668-z} {\bibfield  {journal} {\bibinfo
  {journal} {Nature}\ }\textbf {\bibinfo {volume} {563}},\ \bibinfo {pages}
  {527} (\bibinfo {year} {2018})}\BibitemShut {NoStop}%
\bibitem [{\citenamefont {Pino}\ \emph
  {et~al.}(2021{\natexlab{b}})\citenamefont {Pino}, \citenamefont {Dreiling},
  \citenamefont {Figgatt}, \citenamefont {Gaebler}, \citenamefont {Moses},
  \citenamefont {Allman}, \citenamefont {Baldwin}, \citenamefont {Foss-Feig},
  \citenamefont {Hayes}, \citenamefont {Mayer}, \citenamefont {Ryan-Anderson},\
  and\ \citenamefont {Neyenhuis}}]{Honeywell2021}%
  \BibitemOpen
  \bibfield  {author} {\bibinfo {author} {\bibfnamefont {J.~M.}\ \bibnamefont
  {Pino}}, \bibinfo {author} {\bibfnamefont {J.~M.}\ \bibnamefont {Dreiling}},
  \bibinfo {author} {\bibfnamefont {C.}~\bibnamefont {Figgatt}}, \bibinfo
  {author} {\bibfnamefont {J.~P.}\ \bibnamefont {Gaebler}}, \bibinfo {author}
  {\bibfnamefont {S.~A.}\ \bibnamefont {Moses}}, \bibinfo {author}
  {\bibfnamefont {M.~S.}\ \bibnamefont {Allman}}, \bibinfo {author}
  {\bibfnamefont {C.~H.}\ \bibnamefont {Baldwin}}, \bibinfo {author}
  {\bibfnamefont {M.}~\bibnamefont {Foss-Feig}}, \bibinfo {author}
  {\bibfnamefont {D.}~\bibnamefont {Hayes}}, \bibinfo {author} {\bibfnamefont
  {K.}~\bibnamefont {Mayer}}, \bibinfo {author} {\bibfnamefont
  {C.}~\bibnamefont {Ryan-Anderson}}, \ and\ \bibinfo {author} {\bibfnamefont
  {B.}~\bibnamefont {Neyenhuis}},\ }\href {\doibase 10.1038/s41586-021-03318-4}
  {\bibfield  {journal} {\bibinfo  {journal} {Nature}\ }\textbf {\bibinfo
  {volume} {592}},\ \bibinfo {pages} {209} (\bibinfo {year}
  {2021}{\natexlab{b}})}\BibitemShut {NoStop}%
\bibitem [{\citenamefont {Hilder}\ \emph {et~al.}(2021)\citenamefont {Hilder},
  \citenamefont {Pijn}, \citenamefont {Onishchenko}, \citenamefont {Stahl},
  \citenamefont {Orth}, \citenamefont {Lekitsch}, \citenamefont
  {Rodriguez-Blanco}, \citenamefont {M{\"u}ller}, \citenamefont
  {Schmidt-Kaler},\ and\ \citenamefont {Poschinger}}]{Mainz2021}%
  \BibitemOpen
  \bibfield  {author} {\bibinfo {author} {\bibfnamefont {J.}~\bibnamefont
  {Hilder}}, \bibinfo {author} {\bibfnamefont {D.}~\bibnamefont {Pijn}},
  \bibinfo {author} {\bibfnamefont {O.}~\bibnamefont {Onishchenko}}, \bibinfo
  {author} {\bibfnamefont {A.}~\bibnamefont {Stahl}}, \bibinfo {author}
  {\bibfnamefont {M.}~\bibnamefont {Orth}}, \bibinfo {author} {\bibfnamefont
  {B.}~\bibnamefont {Lekitsch}}, \bibinfo {author} {\bibfnamefont
  {A.}~\bibnamefont {Rodriguez-Blanco}}, \bibinfo {author} {\bibfnamefont
  {M.}~\bibnamefont {M{\"u}ller}}, \bibinfo {author} {\bibfnamefont
  {F.}~\bibnamefont {Schmidt-Kaler}}, \ and\ \bibinfo {author} {\bibfnamefont
  {U.}~\bibnamefont {Poschinger}},\ }\href@noop {} {\bibfield  {journal}
  {\bibinfo  {journal} {arXiv preprint arXiv:2107.06368}\ } (\bibinfo {year}
  {2021})}\BibitemShut {NoStop}%
\bibitem [{\citenamefont {Shapira}\ \emph {et~al.}(2018)\citenamefont
  {Shapira}, \citenamefont {Shaniv}, \citenamefont {Manovitz}, \citenamefont
  {Akerman},\ and\ \citenamefont {Ozeri}}]{Shapira2018}%
  \BibitemOpen
  \bibfield  {author} {\bibinfo {author} {\bibfnamefont {Y.}~\bibnamefont
  {Shapira}}, \bibinfo {author} {\bibfnamefont {R.}~\bibnamefont {Shaniv}},
  \bibinfo {author} {\bibfnamefont {T.}~\bibnamefont {Manovitz}}, \bibinfo
  {author} {\bibfnamefont {N.}~\bibnamefont {Akerman}}, \ and\ \bibinfo
  {author} {\bibfnamefont {R.}~\bibnamefont {Ozeri}},\ }\href {\doibase
  10.1103/PhysRevLett.121.180502} {\bibfield  {journal} {\bibinfo  {journal}
  {Phys. Rev. Lett.}\ }\textbf {\bibinfo {volume} {121}},\ \bibinfo {pages}
  {180502} (\bibinfo {year} {2018})}\BibitemShut {NoStop}%
\bibitem [{\citenamefont {Webb}\ \emph {et~al.}(2018)\citenamefont {Webb},
  \citenamefont {Webster}, \citenamefont {Collingbourne}, \citenamefont
  {Bretaud}, \citenamefont {Lawrence}, \citenamefont {Weidt}, \citenamefont
  {Mintert},\ and\ \citenamefont {Hensinger}}]{webb2018resilient}%
  \BibitemOpen
  \bibfield  {author} {\bibinfo {author} {\bibfnamefont {A.~E.}\ \bibnamefont
  {Webb}}, \bibinfo {author} {\bibfnamefont {S.~C.}\ \bibnamefont {Webster}},
  \bibinfo {author} {\bibfnamefont {S.}~\bibnamefont {Collingbourne}}, \bibinfo
  {author} {\bibfnamefont {D.}~\bibnamefont {Bretaud}}, \bibinfo {author}
  {\bibfnamefont {A.~M.}\ \bibnamefont {Lawrence}}, \bibinfo {author}
  {\bibfnamefont {S.}~\bibnamefont {Weidt}}, \bibinfo {author} {\bibfnamefont
  {F.}~\bibnamefont {Mintert}}, \ and\ \bibinfo {author} {\bibfnamefont
  {W.~K.}\ \bibnamefont {Hensinger}},\ }\href@noop {} {\bibfield  {journal}
  {\bibinfo  {journal} {Physical review letters}\ }\textbf {\bibinfo {volume}
  {121}},\ \bibinfo {pages} {180501} (\bibinfo {year} {2018})}\BibitemShut
  {NoStop}%
\bibitem [{\citenamefont {Zarantonello}\ \emph {et~al.}(2019)\citenamefont
  {Zarantonello}, \citenamefont {Hahn}, \citenamefont {Morgner}, \citenamefont
  {Schulte}, \citenamefont {Bautista-Salvador}, \citenamefont {Werner},
  \citenamefont {Hammerer},\ and\ \citenamefont
  {Ospelkaus}}]{zarantonello2019robust}%
  \BibitemOpen
  \bibfield  {author} {\bibinfo {author} {\bibfnamefont {G.}~\bibnamefont
  {Zarantonello}}, \bibinfo {author} {\bibfnamefont {H.}~\bibnamefont {Hahn}},
  \bibinfo {author} {\bibfnamefont {J.}~\bibnamefont {Morgner}}, \bibinfo
  {author} {\bibfnamefont {M.}~\bibnamefont {Schulte}}, \bibinfo {author}
  {\bibfnamefont {A.}~\bibnamefont {Bautista-Salvador}}, \bibinfo {author}
  {\bibfnamefont {R.}~\bibnamefont {Werner}}, \bibinfo {author} {\bibfnamefont
  {K.}~\bibnamefont {Hammerer}}, \ and\ \bibinfo {author} {\bibfnamefont
  {C.}~\bibnamefont {Ospelkaus}},\ }\href@noop {} {\bibfield  {journal}
  {\bibinfo  {journal} {Physical review letters}\ }\textbf {\bibinfo {volume}
  {123}},\ \bibinfo {pages} {260503} (\bibinfo {year} {2019})}\BibitemShut
  {NoStop}%
\bibitem [{\citenamefont {Manovitz}\ \emph {et~al.}(2017)\citenamefont
  {Manovitz}, \citenamefont {Rotem}, \citenamefont {Shaniv}, \citenamefont
  {Cohen}, \citenamefont {Shapira}, \citenamefont {Akerman}, \citenamefont
  {Retzker},\ and\ \citenamefont {Ozeri}}]{Manovitz2017}%
  \BibitemOpen
  \bibfield  {author} {\bibinfo {author} {\bibfnamefont {T.}~\bibnamefont
  {Manovitz}}, \bibinfo {author} {\bibfnamefont {A.}~\bibnamefont {Rotem}},
  \bibinfo {author} {\bibfnamefont {R.}~\bibnamefont {Shaniv}}, \bibinfo
  {author} {\bibfnamefont {I.}~\bibnamefont {Cohen}}, \bibinfo {author}
  {\bibfnamefont {Y.}~\bibnamefont {Shapira}}, \bibinfo {author} {\bibfnamefont
  {N.}~\bibnamefont {Akerman}}, \bibinfo {author} {\bibfnamefont
  {A.}~\bibnamefont {Retzker}}, \ and\ \bibinfo {author} {\bibfnamefont
  {R.}~\bibnamefont {Ozeri}},\ }\href {\doibase 10.1103/PhysRevLett.119.220505}
  {\bibfield  {journal} {\bibinfo  {journal} {Phys. Rev. Lett.}\ }\textbf
  {\bibinfo {volume} {119}},\ \bibinfo {pages} {220505} (\bibinfo {year}
  {2017})}\BibitemShut {NoStop}%
\bibitem [{\citenamefont {Akerman}\ \emph {et~al.}(2011)\citenamefont
  {Akerman}, \citenamefont {Glickman}, \citenamefont {Kotler}, \citenamefont
  {Keselman},\ and\ \citenamefont {Ozeri}}]{Akerman2011}%
  \BibitemOpen
  \bibfield  {author} {\bibinfo {author} {\bibfnamefont {N.}~\bibnamefont
  {Akerman}}, \bibinfo {author} {\bibfnamefont {Y.}~\bibnamefont {Glickman}},
  \bibinfo {author} {\bibfnamefont {S.}~\bibnamefont {Kotler}}, \bibinfo
  {author} {\bibfnamefont {A.}~\bibnamefont {Keselman}}, \ and\ \bibinfo
  {author} {\bibfnamefont {R.}~\bibnamefont {Ozeri}},\ }\href {\doibase
  10.1007/s00340-011-4807-6} {\bibfield  {journal} {\bibinfo  {journal}
  {Applied Physics B}\ }\textbf {\bibinfo {volume} {107}},\ \bibinfo {pages}
  {1167} (\bibinfo {year} {2011})}\BibitemShut {NoStop}%
\bibitem [{\citenamefont {Akerman}(2012)}]{Akerman:2012}%
  \BibitemOpen
  \bibfield  {author} {\bibinfo {author} {\bibfnamefont {N.}~\bibnamefont
  {Akerman}},\ }\emph {\bibinfo {title} {Trapped ions and free photons}},\
  \href@noop {} {Ph.D. thesis},\ \bibinfo  {school} {Weizmann Institute of
  Science} (\bibinfo {year} {2012})\BibitemShut {NoStop}%
\bibitem [{\citenamefont {Peleg}\ \emph {et~al.}(2019)\citenamefont {Peleg},
  \citenamefont {Akerman}, \citenamefont {Manovitz}, \citenamefont {Alon},\
  and\ \citenamefont {Ozeri}}]{Peleg2019}%
  \BibitemOpen
  \bibfield  {author} {\bibinfo {author} {\bibfnamefont {L.}~\bibnamefont
  {Peleg}}, \bibinfo {author} {\bibfnamefont {N.}~\bibnamefont {Akerman}},
  \bibinfo {author} {\bibfnamefont {T.}~\bibnamefont {Manovitz}}, \bibinfo
  {author} {\bibfnamefont {M.}~\bibnamefont {Alon}}, \ and\ \bibinfo {author}
  {\bibfnamefont {R.}~\bibnamefont {Ozeri}},\ }\href@noop {} {\bibfield
  {journal} {\bibinfo  {journal} {arXiv preprint arXiv:1905.05065}\ } (\bibinfo
  {year} {2019})}\BibitemShut {NoStop}%
\bibitem [{\citenamefont {Lechner}\ \emph {et~al.}(2016)\citenamefont
  {Lechner}, \citenamefont {Maier}, \citenamefont {Hempel}, \citenamefont
  {Jurcevic}, \citenamefont {Lanyon}, \citenamefont {Monz}, \citenamefont
  {Brownnutt}, \citenamefont {Blatt},\ and\ \citenamefont
  {Roos}}]{lechner2016eit}%
  \BibitemOpen
  \bibfield  {author} {\bibinfo {author} {\bibfnamefont {R.}~\bibnamefont
  {Lechner}}, \bibinfo {author} {\bibfnamefont {C.}~\bibnamefont {Maier}},
  \bibinfo {author} {\bibfnamefont {C.}~\bibnamefont {Hempel}}, \bibinfo
  {author} {\bibfnamefont {P.}~\bibnamefont {Jurcevic}}, \bibinfo {author}
  {\bibfnamefont {B.~P.}\ \bibnamefont {Lanyon}}, \bibinfo {author}
  {\bibfnamefont {T.}~\bibnamefont {Monz}}, \bibinfo {author} {\bibfnamefont
  {M.}~\bibnamefont {Brownnutt}}, \bibinfo {author} {\bibfnamefont
  {R.}~\bibnamefont {Blatt}}, \ and\ \bibinfo {author} {\bibfnamefont {C.~F.}\
  \bibnamefont {Roos}},\ }\href {\doibase 10.1103/PhysRevA.93.053401}
  {\bibfield  {journal} {\bibinfo  {journal} {Phys. Rev. A}\ }\textbf {\bibinfo
  {volume} {93}},\ \bibinfo {pages} {053401} (\bibinfo {year}
  {2016})}\BibitemShut {NoStop}%
\bibitem [{\citenamefont {James}(2000)}]{james2000quantum}%
  \BibitemOpen
  \bibfield  {author} {\bibinfo {author} {\bibfnamefont {D.~F.}\ \bibnamefont
  {James}},\ }\href@noop {} {\bibfield  {journal} {\bibinfo  {journal} {Quantum
  Computation and Quantum Information Theory: Reprint Volume with Introductory
  Notes for ISI TMR Network School, 12-23 July 1999, Villa Gualino, Torino,
  Italy}\ }\textbf {\bibinfo {volume} {66}},\ \bibinfo {pages} {345} (\bibinfo
  {year} {2000})}\BibitemShut {NoStop}%
\bibitem [{\citenamefont {Pogorelov}\ \emph
  {et~al.}(2021{\natexlab{b}})\citenamefont {Pogorelov}, \citenamefont
  {Feldker}, \citenamefont {Marciniak}, \citenamefont {Postler}, \citenamefont
  {Jacob}, \citenamefont {Krieglsteiner}, \citenamefont {Podlesnic},
  \citenamefont {Meth}, \citenamefont {Negnevitsky}, \citenamefont {Stadler}
  \emph {et~al.}}]{pogorelov2021compact}%
  \BibitemOpen
  \bibfield  {author} {\bibinfo {author} {\bibfnamefont {I.}~\bibnamefont
  {Pogorelov}}, \bibinfo {author} {\bibfnamefont {T.}~\bibnamefont {Feldker}},
  \bibinfo {author} {\bibfnamefont {C.~D.}\ \bibnamefont {Marciniak}}, \bibinfo
  {author} {\bibfnamefont {L.}~\bibnamefont {Postler}}, \bibinfo {author}
  {\bibfnamefont {G.}~\bibnamefont {Jacob}}, \bibinfo {author} {\bibfnamefont
  {O.}~\bibnamefont {Krieglsteiner}}, \bibinfo {author} {\bibfnamefont
  {V.}~\bibnamefont {Podlesnic}}, \bibinfo {author} {\bibfnamefont
  {M.}~\bibnamefont {Meth}}, \bibinfo {author} {\bibfnamefont {V.}~\bibnamefont
  {Negnevitsky}}, \bibinfo {author} {\bibfnamefont {M.}~\bibnamefont
  {Stadler}},  \emph {et~al.},\ }\href@noop {} {\bibfield  {journal} {\bibinfo
  {journal} {PRX Quantum}\ }\textbf {\bibinfo {volume} {2}},\ \bibinfo {pages}
  {020343} (\bibinfo {year} {2021}{\natexlab{b}})}\BibitemShut {NoStop}%
\bibitem [{\citenamefont {Mintert}\ and\ \citenamefont
  {Wunderlich}(2001)}]{mintert2001ion}%
  \BibitemOpen
  \bibfield  {author} {\bibinfo {author} {\bibfnamefont {F.}~\bibnamefont
  {Mintert}}\ and\ \bibinfo {author} {\bibfnamefont {C.}~\bibnamefont
  {Wunderlich}},\ }\href {\doibase 10.1103/PhysRevLett.87.257904} {\bibfield
  {journal} {\bibinfo  {journal} {Phys. Rev. Lett.}\ }\textbf {\bibinfo
  {volume} {87}},\ \bibinfo {pages} {257904} (\bibinfo {year}
  {2001})}\BibitemShut {NoStop}%
\bibitem [{\citenamefont {Johanning}\ \emph {et~al.}(2009)\citenamefont
  {Johanning}, \citenamefont {Braun}, \citenamefont {Timoney}, \citenamefont
  {Elman}, \citenamefont {Neuhauser},\ and\ \citenamefont
  {Wunderlich}}]{johanning2009individual}%
  \BibitemOpen
  \bibfield  {author} {\bibinfo {author} {\bibfnamefont {M.}~\bibnamefont
  {Johanning}}, \bibinfo {author} {\bibfnamefont {A.}~\bibnamefont {Braun}},
  \bibinfo {author} {\bibfnamefont {N.}~\bibnamefont {Timoney}}, \bibinfo
  {author} {\bibfnamefont {V.}~\bibnamefont {Elman}}, \bibinfo {author}
  {\bibfnamefont {W.}~\bibnamefont {Neuhauser}}, \ and\ \bibinfo {author}
  {\bibfnamefont {C.}~\bibnamefont {Wunderlich}},\ }\href {\doibase
  10.1103/PhysRevLett.102.073004} {\bibfield  {journal} {\bibinfo  {journal}
  {Phys. Rev. Lett.}\ }\textbf {\bibinfo {volume} {102}},\ \bibinfo {pages}
  {073004} (\bibinfo {year} {2009})}\BibitemShut {NoStop}%
\bibitem [{\citenamefont {Timoney}\ \emph {et~al.}(2011)\citenamefont
  {Timoney}, \citenamefont {Baumgart}, \citenamefont {Johanning}, \citenamefont
  {Var{\'o}n}, \citenamefont {Plenio}, \citenamefont {Retzker},\ and\
  \citenamefont {Wunderlich}}]{timoney2011quantum}%
  \BibitemOpen
  \bibfield  {author} {\bibinfo {author} {\bibfnamefont {N.}~\bibnamefont
  {Timoney}}, \bibinfo {author} {\bibfnamefont {I.}~\bibnamefont {Baumgart}},
  \bibinfo {author} {\bibfnamefont {M.}~\bibnamefont {Johanning}}, \bibinfo
  {author} {\bibfnamefont {A.}~\bibnamefont {Var{\'o}n}}, \bibinfo {author}
  {\bibfnamefont {M.~B.}\ \bibnamefont {Plenio}}, \bibinfo {author}
  {\bibfnamefont {A.}~\bibnamefont {Retzker}}, \ and\ \bibinfo {author}
  {\bibfnamefont {C.}~\bibnamefont {Wunderlich}},\ }\href@noop {} {\bibfield
  {journal} {\bibinfo  {journal} {Nature}\ }\textbf {\bibinfo {volume} {476}},\
  \bibinfo {pages} {185} (\bibinfo {year} {2011})}\BibitemShut {NoStop}%
\bibitem [{\citenamefont {Navon}\ \emph {et~al.}(2013)\citenamefont {Navon},
  \citenamefont {Kotler}, \citenamefont {Akerman}, \citenamefont {Glickman},
  \citenamefont {Almog},\ and\ \citenamefont {Ozeri}}]{navon2013addressing}%
  \BibitemOpen
  \bibfield  {author} {\bibinfo {author} {\bibfnamefont {N.}~\bibnamefont
  {Navon}}, \bibinfo {author} {\bibfnamefont {S.}~\bibnamefont {Kotler}},
  \bibinfo {author} {\bibfnamefont {N.}~\bibnamefont {Akerman}}, \bibinfo
  {author} {\bibfnamefont {Y.}~\bibnamefont {Glickman}}, \bibinfo {author}
  {\bibfnamefont {I.}~\bibnamefont {Almog}}, \ and\ \bibinfo {author}
  {\bibfnamefont {R.}~\bibnamefont {Ozeri}},\ }\href {\doibase
  10.1103/PhysRevLett.111.073001} {\bibfield  {journal} {\bibinfo  {journal}
  {Phys. Rev. Lett.}\ }\textbf {\bibinfo {volume} {111}},\ \bibinfo {pages}
  {073001} (\bibinfo {year} {2013})}\BibitemShut {NoStop}%
\bibitem [{\citenamefont {Manovitz}(2016)}]{Manovitz:2016}%
  \BibitemOpen
  \bibfield  {author} {\bibinfo {author} {\bibfnamefont {T.}~\bibnamefont
  {Manovitz}},\ }\emph {\bibinfo {title} {Individual addressing and imaging of
  ions in a {P}aul trap}},\ \href@noop {} {Master's thesis},\ \bibinfo
  {school} {Weizmann Institute of Science} (\bibinfo {year} {2016})\BibitemShut
  {NoStop}%
\bibitem [{\citenamefont {S\o{}rensen}\ and\ \citenamefont
  {M\o{}lmer}(1999)}]{sorensen1999quantum}%
  \BibitemOpen
  \bibfield  {author} {\bibinfo {author} {\bibfnamefont {A.}~\bibnamefont
  {S\o{}rensen}}\ and\ \bibinfo {author} {\bibfnamefont {K.}~\bibnamefont
  {M\o{}lmer}},\ }\href {\doibase 10.1103/PhysRevLett.82.1971} {\bibfield
  {journal} {\bibinfo  {journal} {Phys. Rev. Lett.}\ }\textbf {\bibinfo
  {volume} {82}},\ \bibinfo {pages} {1971} (\bibinfo {year}
  {1999})}\BibitemShut {NoStop}%
\bibitem [{\citenamefont {S\o{}rensen}\ and\ \citenamefont
  {M\o{}lmer}(2000)}]{sorensen2000entanglement}%
  \BibitemOpen
  \bibfield  {author} {\bibinfo {author} {\bibfnamefont {A.}~\bibnamefont
  {S\o{}rensen}}\ and\ \bibinfo {author} {\bibfnamefont {K.}~\bibnamefont
  {M\o{}lmer}},\ }\href {\doibase 10.1103/PhysRevA.62.022311} {\bibfield
  {journal} {\bibinfo  {journal} {Phys. Rev. A}\ }\textbf {\bibinfo {volume}
  {62}},\ \bibinfo {pages} {022311} (\bibinfo {year} {2000})}\BibitemShut
  {NoStop}%
\bibitem [{\citenamefont {Gaebler}\ \emph {et~al.}(2016)\citenamefont
  {Gaebler}, \citenamefont {Tan}, \citenamefont {Lin}, \citenamefont {Wan},
  \citenamefont {Bowler}, \citenamefont {Keith}, \citenamefont {Glancy},
  \citenamefont {Coakley}, \citenamefont {Knill}, \citenamefont {Leibfried}
  \emph {et~al.}}]{gaebler2016high}%
  \BibitemOpen
  \bibfield  {author} {\bibinfo {author} {\bibfnamefont {J.~P.}\ \bibnamefont
  {Gaebler}}, \bibinfo {author} {\bibfnamefont {T.~R.}\ \bibnamefont {Tan}},
  \bibinfo {author} {\bibfnamefont {Y.}~\bibnamefont {Lin}}, \bibinfo {author}
  {\bibfnamefont {Y.}~\bibnamefont {Wan}}, \bibinfo {author} {\bibfnamefont
  {R.}~\bibnamefont {Bowler}}, \bibinfo {author} {\bibfnamefont {A.~C.}\
  \bibnamefont {Keith}}, \bibinfo {author} {\bibfnamefont {S.}~\bibnamefont
  {Glancy}}, \bibinfo {author} {\bibfnamefont {K.}~\bibnamefont {Coakley}},
  \bibinfo {author} {\bibfnamefont {E.}~\bibnamefont {Knill}}, \bibinfo
  {author} {\bibfnamefont {D.}~\bibnamefont {Leibfried}},  \emph {et~al.},\
  }\href@noop {} {\bibfield  {journal} {\bibinfo  {journal} {Physical review
  letters}\ }\textbf {\bibinfo {volume} {117}},\ \bibinfo {pages} {060505}
  (\bibinfo {year} {2016})}\BibitemShut {NoStop}%
\bibitem [{\citenamefont {Ballance}\ \emph {et~al.}(2016)\citenamefont
  {Ballance}, \citenamefont {Harty}, \citenamefont {Linke}, \citenamefont
  {Sepiol},\ and\ \citenamefont {Lucas}}]{ballance2016high}%
  \BibitemOpen
  \bibfield  {author} {\bibinfo {author} {\bibfnamefont {C.}~\bibnamefont
  {Ballance}}, \bibinfo {author} {\bibfnamefont {T.}~\bibnamefont {Harty}},
  \bibinfo {author} {\bibfnamefont {N.}~\bibnamefont {Linke}}, \bibinfo
  {author} {\bibfnamefont {M.}~\bibnamefont {Sepiol}}, \ and\ \bibinfo {author}
  {\bibfnamefont {D.}~\bibnamefont {Lucas}},\ }\href@noop {} {\bibfield
  {journal} {\bibinfo  {journal} {Physical review letters}\ }\textbf {\bibinfo
  {volume} {117}},\ \bibinfo {pages} {060504} (\bibinfo {year}
  {2016})}\BibitemShut {NoStop}%
\bibitem [{\citenamefont {Srinivas}\ \emph {et~al.}(2021)\citenamefont
  {Srinivas}, \citenamefont {Burd}, \citenamefont {Knaack}, \citenamefont
  {Sutherland}, \citenamefont {Kwiatkowski}, \citenamefont {Glancy},
  \citenamefont {Knill}, \citenamefont {Wineland}, \citenamefont {Leibfried},
  \citenamefont {Wilson} \emph {et~al.}}]{srinivas2021high}%
  \BibitemOpen
  \bibfield  {author} {\bibinfo {author} {\bibfnamefont {R.}~\bibnamefont
  {Srinivas}}, \bibinfo {author} {\bibfnamefont {S.}~\bibnamefont {Burd}},
  \bibinfo {author} {\bibfnamefont {H.}~\bibnamefont {Knaack}}, \bibinfo
  {author} {\bibfnamefont {R.}~\bibnamefont {Sutherland}}, \bibinfo {author}
  {\bibfnamefont {A.}~\bibnamefont {Kwiatkowski}}, \bibinfo {author}
  {\bibfnamefont {S.}~\bibnamefont {Glancy}}, \bibinfo {author} {\bibfnamefont
  {E.}~\bibnamefont {Knill}}, \bibinfo {author} {\bibfnamefont
  {D.}~\bibnamefont {Wineland}}, \bibinfo {author} {\bibfnamefont
  {D.}~\bibnamefont {Leibfried}}, \bibinfo {author} {\bibfnamefont
  {A.}~\bibnamefont {Wilson}},  \emph {et~al.},\ }\href@noop {} {\bibfield
  {journal} {\bibinfo  {journal} {arXiv preprint arXiv:2102.12533}\ } (\bibinfo
  {year} {2021})}\BibitemShut {NoStop}%
\bibitem [{\citenamefont {Clark}\ \emph {et~al.}(2021)\citenamefont {Clark},
  \citenamefont {Tinkey}, \citenamefont {Sawyer}, \citenamefont {Meier},
  \citenamefont {Burkhardt}, \citenamefont {Seck}, \citenamefont {Shappert},
  \citenamefont {Guise}, \citenamefont {Volin}, \citenamefont {Fallek} \emph
  {et~al.}}]{clark2021high}%
  \BibitemOpen
  \bibfield  {author} {\bibinfo {author} {\bibfnamefont {C.~R.}\ \bibnamefont
  {Clark}}, \bibinfo {author} {\bibfnamefont {H.~N.}\ \bibnamefont {Tinkey}},
  \bibinfo {author} {\bibfnamefont {B.~C.}\ \bibnamefont {Sawyer}}, \bibinfo
  {author} {\bibfnamefont {A.~M.}\ \bibnamefont {Meier}}, \bibinfo {author}
  {\bibfnamefont {K.~A.}\ \bibnamefont {Burkhardt}}, \bibinfo {author}
  {\bibfnamefont {C.~M.}\ \bibnamefont {Seck}}, \bibinfo {author}
  {\bibfnamefont {C.~M.}\ \bibnamefont {Shappert}}, \bibinfo {author}
  {\bibfnamefont {N.~D.}\ \bibnamefont {Guise}}, \bibinfo {author}
  {\bibfnamefont {C.~E.}\ \bibnamefont {Volin}}, \bibinfo {author}
  {\bibfnamefont {S.~D.}\ \bibnamefont {Fallek}},  \emph {et~al.},\ }\href@noop
  {} {\bibfield  {journal} {\bibinfo  {journal} {arXiv preprint
  arXiv:2105.05828}\ } (\bibinfo {year} {2021})}\BibitemShut {NoStop}%
\bibitem [{\citenamefont {Shapira}\ \emph {et~al.}(2020)\citenamefont
  {Shapira}, \citenamefont {Shaniv}, \citenamefont {Manovitz}, \citenamefont
  {Akerman}, \citenamefont {Peleg}, \citenamefont {Gazit}, \citenamefont
  {Ozeri},\ and\ \citenamefont {Stern}}]{Shapira2020}%
  \BibitemOpen
  \bibfield  {author} {\bibinfo {author} {\bibfnamefont {Y.}~\bibnamefont
  {Shapira}}, \bibinfo {author} {\bibfnamefont {R.}~\bibnamefont {Shaniv}},
  \bibinfo {author} {\bibfnamefont {T.}~\bibnamefont {Manovitz}}, \bibinfo
  {author} {\bibfnamefont {N.}~\bibnamefont {Akerman}}, \bibinfo {author}
  {\bibfnamefont {L.}~\bibnamefont {Peleg}}, \bibinfo {author} {\bibfnamefont
  {L.}~\bibnamefont {Gazit}}, \bibinfo {author} {\bibfnamefont
  {R.}~\bibnamefont {Ozeri}}, \ and\ \bibinfo {author} {\bibfnamefont
  {A.}~\bibnamefont {Stern}},\ }\href {\doibase 10.1103/PhysRevA.101.032330}
  {\bibfield  {journal} {\bibinfo  {journal} {Phys. Rev. A}\ }\textbf {\bibinfo
  {volume} {101}},\ \bibinfo {pages} {032330} (\bibinfo {year}
  {2020})}\BibitemShut {NoStop}%
\bibitem [{\citenamefont {Figgatt}\ \emph {et~al.}(2019)\citenamefont
  {Figgatt}, \citenamefont {Ostrander}, \citenamefont {Linke}, \citenamefont
  {Landsman}, \citenamefont {Zhu}, \citenamefont {Maslov},\ and\ \citenamefont
  {Monroe}}]{figgatt2019parallel}%
  \BibitemOpen
  \bibfield  {author} {\bibinfo {author} {\bibfnamefont {C.}~\bibnamefont
  {Figgatt}}, \bibinfo {author} {\bibfnamefont {A.}~\bibnamefont {Ostrander}},
  \bibinfo {author} {\bibfnamefont {N.~M.}\ \bibnamefont {Linke}}, \bibinfo
  {author} {\bibfnamefont {K.~A.}\ \bibnamefont {Landsman}}, \bibinfo {author}
  {\bibfnamefont {D.}~\bibnamefont {Zhu}}, \bibinfo {author} {\bibfnamefont
  {D.}~\bibnamefont {Maslov}}, \ and\ \bibinfo {author} {\bibfnamefont
  {C.}~\bibnamefont {Monroe}},\ }\href@noop {} {\bibfield  {journal} {\bibinfo
  {journal} {Nature}\ }\textbf {\bibinfo {volume} {572}},\ \bibinfo {pages}
  {368} (\bibinfo {year} {2019})}\BibitemShut {NoStop}%
\bibitem [{\citenamefont {Aharonov}\ and\ \citenamefont
  {Ben-Or}(1997)}]{aharonov1997fault}%
  \BibitemOpen
  \bibfield  {author} {\bibinfo {author} {\bibfnamefont {D.}~\bibnamefont
  {Aharonov}}\ and\ \bibinfo {author} {\bibfnamefont {M.}~\bibnamefont
  {Ben-Or}},\ }in\ \href@noop {} {\emph {\bibinfo {booktitle} {Proceedings of
  the twenty-ninth annual ACM symposium on Theory of computing}}}\ (\bibinfo
  {year} {1997})\ pp.\ \bibinfo {pages} {176--188}\BibitemShut {NoStop}%
\bibitem [{\citenamefont {Knill}\ and\ \citenamefont
  {Laflamme}(1997)}]{knill1997theory}%
  \BibitemOpen
  \bibfield  {author} {\bibinfo {author} {\bibfnamefont {E.}~\bibnamefont
  {Knill}}\ and\ \bibinfo {author} {\bibfnamefont {R.}~\bibnamefont
  {Laflamme}},\ }\href@noop {} {\bibfield  {journal} {\bibinfo  {journal}
  {Physical Review A}\ }\textbf {\bibinfo {volume} {55}},\ \bibinfo {pages}
  {900} (\bibinfo {year} {1997})}\BibitemShut {NoStop}%
\bibitem [{\citenamefont {Gazit}(2020)}]{Gazit:2020}%
  \BibitemOpen
  \bibfield  {author} {\bibinfo {author} {\bibfnamefont {L.}~\bibnamefont
  {Gazit}},\ }\emph {\bibinfo {title} {Quantum feedback on trapped ions}},\
  \href
  {https://www.weizmann.ac.il/complex/ozeri/sites/complex.ozeri/files/uploads/ths241220152915_msc2_7684.pdf}
  {Master's thesis},\ \bibinfo  {school} {Weizmann Institute of Science}
  (\bibinfo {year} {2020})\BibitemShut {NoStop}%
\bibitem [{\citenamefont {Carr}\ and\ \citenamefont
  {Purcell}(1954)}]{carr1954effects}%
  \BibitemOpen
  \bibfield  {author} {\bibinfo {author} {\bibfnamefont {H.~Y.}\ \bibnamefont
  {Carr}}\ and\ \bibinfo {author} {\bibfnamefont {E.~M.}\ \bibnamefont
  {Purcell}},\ }\href@noop {} {\bibfield  {journal} {\bibinfo  {journal}
  {Physical review}\ }\textbf {\bibinfo {volume} {94}},\ \bibinfo {pages} {630}
  (\bibinfo {year} {1954})}\BibitemShut {NoStop}%
\bibitem [{\citenamefont {Meiboom}\ and\ \citenamefont
  {Gill}(1958)}]{meiboom1958modified}%
  \BibitemOpen
  \bibfield  {author} {\bibinfo {author} {\bibfnamefont {S.}~\bibnamefont
  {Meiboom}}\ and\ \bibinfo {author} {\bibfnamefont {D.}~\bibnamefont {Gill}},\
  }\href@noop {} {\bibfield  {journal} {\bibinfo  {journal} {Review of
  scientific instruments}\ }\textbf {\bibinfo {volume} {29}},\ \bibinfo {pages}
  {688} (\bibinfo {year} {1958})}\BibitemShut {NoStop}%
\end{thebibliography}%
